\documentclass[12pt]{article}
\usepackage{amsmath} 
\usepackage{amssymb}
\input epsf.tex    
\def\beq{\begin{equation}}
\def\eeq{\end{equation}}
\def\beqa{\begin{eqnarray}}
\def\eeqa{\end{eqnarray}}

\def\Dsl{\,\raise.15ex\hbox{/}\mkern-13.5mu D} 

\begin{document}
\bigskip\begin{titlepage}
\begin{flushright}
UUITP-19/04\\
\end{flushright}
\vspace{1cm}
\begin{center}
{\Large\bf Lectures on string theory and cosmology\\}
\end{center}
\vspace{3mm}
\begin{center}
{\large
Ulf H.\ Danielsson} \\
\vspace{5mm}
Institutionen f\"or Teoretisk Fysik, Box 803, SE-751 08
Uppsala, Sweden
\vspace{5mm}
{\tt
ulf@teorfys.uu.se \\
}
\end{center}
\vspace{5mm}
\begin{center}
{\large\bf Abstract}
\end{center}
\noindent
In these lectures I review recent attempts to apply string theory to cosmology, including string cosmology and
various models of brane cosmology. In addition, the review includes an introduction to inflation as well as a
discussion of transplanckian signatures.  I also provide a critical discussion of the possible role of holography.
The material is based on lectures given in January 2004 at the RTN String School in Barcelona, but also contain some additional material.
\vfill
\begin{flushleft}
August 2004
\end{flushleft}
\end{titlepage}

\section{Introduction}

\bigskip

String theory has long been viewed as an enterprise of little interest for
experiments and observations. The energy scales usually considered to be
relevant for strings are many orders of magnitude higher than what in the
foreseeable future will be experimentally accessible. There are even some
physicists who claim that the realm of string theory forever will be beyond
the grasp of experimental science. Luckily, there are promising signs that the
situation is about to change. Recent developments show that string theory can
become accessible to observations much sooner than most people have ever
hoped. The new player in the game is cosmology. For a long time an inexact
patchwork of educated guesses and order of magnitude estimates, cosmology has
developed into an exact science with a fruitful and rapid interaction between
observations and theory. Much of the progress is based on the ever more
precise observations of the CMBR, and measurements of how the expansion of the
universe has changed with time. Thanks to these new observations it is now
generally believed that the large scale structure of the universe can be
traced back to microscopical physics near the Big Bang. In this way the
universe works like a gigantic accelerator allowing us to study physics at the
very highest energy scales, possibly even scales relevant for strings.

In the meantime, string theory has reached a maturity which allows for the
formulation of realistic cosmological models. For a long time string theory
focused on the physics of the very smallest scales. The problems, which were
addressed, concerned the unification of forces, including gravity, and the
compatibility of relativity and quantum mechanics. The idea was that once the
fundamental microscopical laws were found the rest of physics would follow. In
particular, cosmology was thought of as just another application of these
fundamental laws. In later years the perspective has changed. Many now believe
that the physics of the large and the small can not be separated, and that an
understanding of unification not only is necessary for understanding the
origin of the universe, but that an understanding of the origin of the
universe is necessary in order to understand unification. To summarize,
cosmology can be the key to the verification of string theory, and string
theory can be what we need to solve several of the present puzzles in cosmology.

In these lectures I will give a review of recent attempts to connect string
theory with cosmology.\footnote{Another review, which covers similar topics,
is \cite{Quevedo:2002xw}.} Any such attempt must, in one way or the other, be
confronted with inflation, \cite{Guth:1980zm}\cite{Linde:1981mu}%
\cite{Albrecht:1982wi}.\footnote{Other early ideas about inflation include
\cite{Starobinsky:1979ty}\cite{Starobinsky:1980te}\cite{Sato:1980yn}.} That
is, the widely held view that the early universe underwent a period of
exponential expansion. A complete theory of the early universe must either
explain inflation or replace it with something else. This is also true for
string theory, and I will therefore start out with a basic review of inflation
focusing on those aspects useful for a string theorist wishing to enter the
field. For a more complete introduction, and a complete list of references, I
recommend \cite{liddle}. Apart from standard material, I will briefly discuss
the issue of transplanckian signatures. That is, the possibility of finding
observational signatures of stringy or planckian physics in the CMBR.

I will then proceed with a discussion of the relation between string theory
and inflation. Can strings give rise to inflation? I will review two sets of
proposals: string cosmology and brane cosmology. The latter can be divided
into two subproposals: models that generate inflation, and models that try to
do with out inflation. I will also discuss some of the difficulties
encountered in constructing string theories in de Sitter space and briefly
mention some important aspects of recent progress in this area. Finally, I
will discuss the relevance of holography to cosmology, and conclude with some
comments on the anthropic principle.

\bigskip

\section{About inflation}

\bigskip

\subsection{What is the problem?}

\bigskip

The standard Big Bang model suffers from a number of annoying problems. One of
them, \textit{the flatness problem}, concerns the observation that the real
density of the universe, $\rho$, long has been known to be very close to the
critical density $\rho_{c}$. That is, $\Omega=\frac{\rho}{\rho_{c}}$ has been
measured to be close to one. To understand the importance of this, we start
with the Friedmann equation
\begin{equation}
H^{2}=\frac{1}{3M_{4}^{2}}\rho-\frac{k}{a^{2}}, \label{feq}%
\end{equation}
where $M_{4}=1/\sqrt{8\pi G}\sim2\cdot10^{18}GeV$ is the four dimensional
(reduced) Planck mass. Furthermore, $H=\frac{\dot{a}}{a}$ is the Hubble
constant and $a\left(  t\right)  $ the scale factor with the space time metric
on the form%
\begin{equation}
ds^{2}=dt^{2}-a^{2}dS^{2}.
\end{equation}
$dS^{2}$ is the comoving volume element of space with $k=0$, $+1$ and $-1$
corresponding to flat, positively curved and negatively curved spaces
respectively. We then rewrite the Friedmann equation as
\begin{equation}
\Omega-1=\frac{k}{a^{2}H^{2}},
\end{equation}
and note that for any ordinary type of matter, $\frac{1}{a^{2}H^{2}}$ will
\textit{increase} with time. To see this, we use the continuum equation given
by%
\begin{equation}
\dot{\rho}+3H\left(  \rho+p\right)  =0. \label{ceq}%
\end{equation}
Assuming an equation of state of the form%
\begin{equation}
p=w\rho,
\end{equation}
where $w$ is a constant, the continuum equation can be rewritten as%
\begin{equation}
\frac{d\rho}{da}+3\left(  1+w\right)  \frac{\rho}{a}=0, \label{cont}%
\end{equation}
giving rise to
\begin{equation}
\rho\sim a^{-3\left(  1+w\right)  }. \label{rhoa}%
\end{equation}
If we start with $\Omega\sim1$ ($k\sim0$) we have $H\sim1/t^{2}$, and the
Friedmann equation gives $a\sim t^{\frac{2}{3\left(  1+w\right)  }}$. As a
consequence we finally find%
\begin{equation}
\frac{1}{a^{2}H^{2}}\sim t^{2-\frac{4}{3\left(  1+w\right)  }}, \label{1ah}%
\end{equation}
which clearly grows rapidly with time for any $w>-1/3$ -- examples include
pressureless dust with $w=0$ and radiation with $w=1/3$.

From the above one concludes that, unless the universe is \textit{exactly}
flat ($k=0)$ and, as a consequence, has \textit{exactly} $\Omega=1$, $\Omega$
will rapidly evolve away from $\Omega=1$. If one starts with a value
$\Omega<1$, \ the value will decrease towards zero, while if $\Omega>1$ the
value of $\Omega$ will increase, and even diverge, if the expansion stops as
shown in figure 1.%
\begin{figure}
\begin{center}
\centering
\epsfysize=3.5cm
\leavevmode
\epsfbox{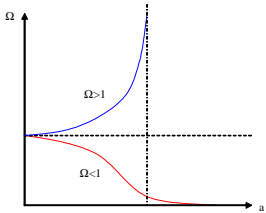}
\end{center}   
\caption[]{\small How $\Omega$ rapidly evolves away from $1$.}
\end{figure}
In order to have a value close to $1$ today, one would therefore expect to
need a value of $\Omega$ even closer to $1$ in the early universe. How close?
Let us assume a radiation dominated universe up to the time $t_{rad}%
\sim300000$ years, and thereafter matter domination. This is roughly the time
when the universe became transparent and the time of origin of the CMBR. We
can then, using (\ref{1ah}) in two steps, estimate the amount of fine tuning
at $t<t_{rad}$ to be%
\begin{equation}
\left\vert \Omega\left(  t\right)  -1\right\vert \sim\frac{t}{t_{rad}}\left(
\frac{t_{rad}}{t_{now}}\right)  ^{2/3}. \label{hminfl}%
\end{equation}
With $t_{now}\sim10^{10}$ years we find a fine tuning of one part in $10^{16}$
one second after the Big Bang, and one part in $10^{60}$ at planckian times
$\sim10^{-44}s$, if the deviation from $\Omega=1$ is to remain small all the
way up to present times. This is the flatness problem. That is, how can
$\Omega$ be so close to one?

Another problem is the \textit{horizon problem}. Regions of the universe, in
particular sources of the CMBR at opposite points of the sky, look very
similar even though, assuming normal radiation dominated expansion in the
early universe, they can not have been in casual contact since the Big Bang.
How is this possible? The problem is illustrated in figure 2.%
\begin{figure}
\begin{center}
\centering
\epsfysize=3.5cm
\leavevmode
\epsfbox{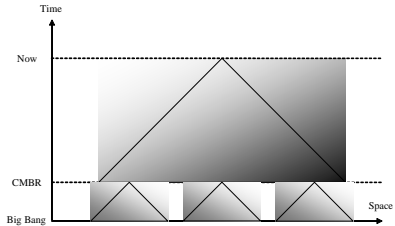}
\end{center}   
\caption[]{\small The horizon problem.}
\end{figure}
In the diagram it can be seen how points at the time when the CMBR was
generated, all visible to us today, have not had time to communicate with each
other. It is difficult to understand how the initial conditions at the Big
Bang could be so extremely fine tuned.

\subsection{Inflation as solution}

\bigskip

A possible way out of the unnatural fine tuning implied by the flatness
problem, would be some kind of mechanism at work in the early universe that
dynamically drives $\Omega$ towards $1$. This is where \textit{inflation}
comes in. Inflation corresponds to a period when $\frac{1}{a^{2}H^{2}}$
actually \textit{decreases}. This is the case for an expanding universe if the
scale factor $a$, that is, the distance between two test objects, increases
faster than the horizon radius $1/H$. In a sense, one can say that the
universe expands faster than the speed of light. In such a universe the
redshift of any given object will increase with time as the object catches up
with the cosmological horizon. Let us see how this works in more detail. A
lightray in the metric%
\begin{equation}
ds^{2}=dt^{2}-a^{2}dx^{2}, \label{plattmet}%
\end{equation}
travels according to%
\begin{equation}
x=\int_{t_{0}}^{t}\frac{dt}{a},
\end{equation}
between time of emission $t_{0}$, and time of observation $t$, where $x$ is
the comoving distance. If we follow a particular object we have $t_{0}%
=t_{0}\left(  t\right)  $, while $x$ is independent of $t$. Differentiating
with respect to $t$, using $dx=0$, we find%
\begin{equation}
\frac{dt}{a\left(  t\right)  }-\frac{dt_{0}}{a\left(  t_{0}\right)
}=0\Longrightarrow\frac{dt_{0}}{dt}=\frac{a\left(  t_{0}\right)  }{a\left(
t\right)  }.
\end{equation}
The redshift of a particular object, as a function of time, is defined by%
\begin{equation}
z\left(  t\right)  =1+\frac{a\left(  t\right)  }{a\left(  t_{0}\left(
t\right)  \right)  }.
\end{equation}
Differentiation with respect to time $t$ gives%
\begin{equation}
\frac{dz}{dt}=\frac{\dot{a}\left(  t\right)  }{a\left(  t_{0}\left(  t\right)
\right)  }-\frac{a\left(  t\right)  }{a\left(  t_{0}\left(  t\right)  \right)
^{2}}\dot{a}\left(  t_{0}\left(  t\right)  \right)  \frac{dt_{0}}{dt}=\frac
{1}{a\left(  t_{0}\right)  }\left(  a\left(  t\right)  H\left(  t\right)
-a\left(  t_{0}\left(  t\right)  \right)  H\left(  t_{0}\right)  \right)  ,
\end{equation}
which is positive if $\frac{d}{dt}\frac{1}{a^{2}H^{2}}<0$, as we set out to
prove. Note that in a universe, which expands in the usual fashion, the
redshift of a given object actually \textit{decreases.}

Faster than light expansion also solves the horizon problem. The reason is, as
explained above, that the expansion rate in a very definite sense is faster
than the speed of light. Objects in causal contact can, through the expansion,
be separated to distances larger than the Hubble radius. Eventually, when
inflation stops, the Hubble radius will start growing faster than the
expansion and the objects will return within their respective horizons. An
observer not taking inflation into account will wrongly conclude that these
objects have never before been in causal contact.

The simplest example of an inflating cosmology is a universe with
$H=\mathrm{const.}$ Such a universe has $a\left(  t\right)  \sim e^{Ht}$ and
is called a $\ $\textit{de Sitter space time.}

\subsection{How do you get inflation?}

We have now seen how inflation solves the problems of the Big Bang model, but
how do we get inflation? The condition for inflation can be written%
\begin{equation}
\frac{d}{dt}\frac{1}{a^{2}H^{2}}=\frac{d}{dt}\frac{1}{\dot{a}^{2}}%
=-\frac{2\ddot{a}}{\dot{a}^{3}}<0,
\end{equation}
or $\ddot{a}>0$ (\textit{if} $\dot{a}>0$), that is, it corresponds to an
\textit{accelerating }expansion. Combining (\ref{feq}) (with $k=0$) and
(\ref{ceq}), one can obtain another Friedmann equation
\begin{equation}
\frac{\overset{\cdot\cdot}{a}}{a}=-\frac{1}{6M_{4}^{2}}\left(  \rho+3p\right)
,
\end{equation}
from which it immediately follows that an accelerated universe requires matter
with negative pressure. Luckily, this can be provided by a scalar field, the
\textit{inflaton}, which possesses a potential energy. Let us investigate this
in more detail.

The Lagrangian for a scalar field is given by%
\begin{equation}
S=\int d^{4}x\sqrt{-g}\left[  \frac{1}{2}\partial^{\mu}\phi\partial_{\mu}%
\phi-V\left(  \phi\right)  \right]  =\int d^{4}xa^{3}\left[  \frac{1}{2}%
\dot{\phi}^{2}-\frac{1}{2a^{2}}\nabla^{2}\phi-V\left(  \phi\right)  \right]  ,
\end{equation}
and the canonical energy momentum tensor is given by%
\begin{equation}
T_{\mu\nu}=\partial_{\mu}\phi\partial_{\nu}\phi-g_{\mu\nu}L.
\end{equation}
In case of a homogenous inflaton field this reduces to an energy density given
by
\begin{equation}
\rho=T_{00}=\dot{\phi}^{2}-\frac{\dot{\phi}^{2}}{2}+V\left(  \phi\right)
=\frac{\dot{\phi}^{2}}{2}+V\left(  \phi\right)  , \label{dphi}%
\end{equation}
and a pressure given by%
\begin{equation}
p=\frac{1}{a^{2}}T_{xx}=\frac{\dot{\phi}^{2}}{2}-V\left(  \phi\right)  .
\label{pphi}%
\end{equation}
Note that $x$ is the comoving coordinate -- hence the rescaling of $T_{xx}$ to
obtain the physical pressure. We also have the equation of motion for the
scalar field given by%
\begin{equation}
\ddot{\phi}+3H\dot{\phi}-\frac{1}{a^{2}}\nabla^{2}\phi+V^{\prime}\left(
\phi\right)  =0. \label{sceq}%
\end{equation}

At this point it is useful to introduce the \textit{slow roll approximation.
}That is, we assume
\begin{equation}
\dot{\phi}^{2}<<V\left(  \phi\right)  ,
\end{equation}
or, in other words, $p\sim-\rho$. We also need to impose $\ddot{\phi}<<V^{\prime
}\left(  \phi\right)  $, and as a consequence we therefore have%
\begin{align}
H^{2}  &  =\frac{1}{3M_{4}^{2}}\left(  \frac{\dot{\phi}^{2}}{2}+V\left(
\phi\right)  \right)  \sim\frac{1}{3M_{4}^{2}}V\left(  \phi\right) \\
3H\dot{\phi}  &  \sim-V^{\prime}\left(  \phi\right)  .
\end{align}
The slow roll conditions are conveniently handled by introducing the
\textit{slow roll parameters}%
\begin{align}
\varepsilon &  =\frac{M_{4}^{2}}{2}\left(  \frac{V^{\prime}}{V}\right)  ^{2}\\
\eta &  =M_{4}^{2}\frac{V^{\prime\prime}}{V}.
\end{align}
It is a useful exercise to verify that the slow roll condition implies that
the slow roll parameters are small. It is also true that inflation implies
that the slow roll parameters are small.

How much inflation do we need to solve the problems of the Big Bang? According
to (\ref{hminfl}) we need a fine tuning of $10^{60}$ at planckian times. If
this is supposed to be achieved through exponential expansion, we must have%
\begin{equation}
a^{2}\sim e^{2Ht}=e^{2N}\sim10^{60}\sim e^{140}.
\end{equation}
That is, we find the required number of e-foldings, $N$, to be around $70$.
This gives a constraint on the potential as follows,%
\begin{equation}
N=\ln\frac{a\left(  t_{e}\right)  }{a\left(  t_{i}\right)  }=\int_{t_{i}%
}^{t_{e}}Hdt=\int_{\phi_{i}}^{\phi_{e}}H\frac{d\phi}{\dot{\phi}}=-\int
_{\phi_{i}}^{\phi_{e}}\frac{3H^{2}}{V^{\prime}}d\phi=-8\pi G\int_{\phi_{i}%
}^{\phi_{e}}\frac{V}{V^{\prime}}d\phi\gtrsim70.
\end{equation}
Using the slow roll parameters we find%
\begin{equation}
N=\frac{1}{\sqrt{2\varepsilon}}\frac{\phi_{i}-\phi_{e}}{M_{4}}\gtrsim70,
\label{n70}%
\end{equation}
and, as a consequence, one concludes that the inflationary potential needs to
be rather flat.

\subsection{A couple of inflationary examples}

Let us now consider a couple of explicit examples. The original works on
inflation assumed potentials with local minima (old inflation), or very flat
maxima (new inflation), in order to keep the inflaton away from the final,
global, minima long enough to get the required number of e-foldings. Later it
was realized that the potential can be of a very simple form. In fact, even a
simple monomial like%
\begin{equation}
V=\lambda M_{4}^{4-\alpha}\phi^{\alpha},
\end{equation}
can do the job. The reason is easy to understand from a quick look at
(\ref{sceq}). The second term in the equation, which is due to the expansion
of the universe, works like a friction term that prevents the inflaton from
rolling down too quickly preventing inflation from taking place. This is
called \textit{chaotic} inflation, \cite{Linde:1983gd}.

For the particular potential above, we can calculate the slow roll parameters
to be%
\begin{equation}
\varepsilon=\frac{\alpha^{2}}{2}\frac{M_{4}^{2}}{\phi^{2}}\qquad\eta
=\alpha\left(  \alpha-1\right)  \frac{M_{4}^{2}}{\phi^{2}}.
\end{equation}
Inflation starts at a large value of $\phi$ and the inflaton then rolls slowly
towards the minimum with increasing $\varepsilon$ and $\left\vert
\eta\right\vert $. Inflation ends when the slow roll conditions no longer
hold, i.e. when $\phi\sim\alpha M_{4}$. The number of e-foldings we obtain
before this happens is given by%
\begin{equation}
N=\frac{1}{M_{4}^{2}}\int_{\phi_{e}}^{\phi_{i}}\frac{\phi}{\alpha}\sim\frac
{1}{2\alpha M_{4}^{2}}\phi_{i}^{2}\Longrightarrow\phi_{i}\sim\sqrt{2\alpha
N}M_{4}\gg M_{4}.
\end{equation}
At the start of inflation the slow roll parameters are given by%
\begin{equation}
\varepsilon\sim\frac{\alpha}{4N}\qquad\eta\sim\frac{\alpha-1}{2N}.
\end{equation}

Another type of potential is%
\begin{equation}
V=V_{0}e^{-\sqrt{\frac{2}{p}}\frac{\phi}{M_{4}}},
\end{equation}
leading to\textit{ power inflation} with $a\sim t^{p}$. In this case the slow
roll parameters are constant and given by%
\begin{equation}
\varepsilon=\frac{1}{p},\qquad\eta=\frac{2}{p}.
\end{equation}
As a result, inflation continues forever with $\phi$ rolling to larger and
larger values. In this case one needs an independent mechanism to end inflation.

\subsection{Quantum fluctuations}

\bigskip

How do we test inflation? The key is structure formation. An important reason
to invoke inflation is to make the universe smooth and flat. In the real
universe, however, there is a large amount of structure. This structure can be
traced back to subtle variations in the matter distribution during the time
when the CMBR was released. A naive application of inflation does, however,
exclude such non-uniformity. So, from where does all the structure come?
Actually, inflation itself supplies the answer provided we take quantum
mechanics into account.

The main insight is that inflation magnifies microscopic quantum fluctuations
into cosmic size, and thereby provides seeds for structure formation. The
details of physics at the highest energy scales is therefore reflected in the
distribution of galaxies and other structures on large scales. The
fluctuations begin their life on the smallest scales and grow larger (in
wavelength) as the universe expands. Eventually they become larger than the
horizon and freeze. That is, different parts of a wave can no longer
communicate with each other since light can not keep up with the expansion of
the universe. This is a consequence of the fact that the scale factor grows
faster than the horizon, which, as we have seen, is a defining property of an
accelerating and inflating universe. At a later time, when inflation stops,
the scale factor will start to grow slower than the horizon and the
fluctuations will eventually come back within the causal horizon. The
fluctuations will then start off acoustic waves in the plasma which will
affect the CMBR. These imprints of the quantum fluctuations can be studied
revealing important clues about physics at extremely high energies in the
early universe.

Let us now investigate in more detail the predictions from inflation. We
assume that the metric as well as the inflaton can be split into a classical
background piece and a piece due to fluctuations according to%
\begin{align}
g_{\mu\nu}  &  =g_{\mu\nu}^{\left(  0\right)  }+h_{\mu\nu}\left(
\tau,\mathbf{x}\right) \\
\phi &  =\phi^{\left(  0\right)  }+\delta\phi\left(  \tau,\mathbf{x}\right)  .
\end{align}
For convenience we have changed coordinates and introduced \textit{conformal
time},\textit{ }$\tau$, such that the metric is given by%
\begin{equation}
ds^{2}=a\left(  \tau\right)  ^{2}\left(  dt^{2}-d\mathbf{x}^{2}\right)  .
\end{equation}
In these coordinates the scalar equation (\ref{sceq}), ignoring the potential
piece, becomes%
\begin{equation}
\delta\phi_{\mathbf{k}}^{\prime\prime}+2\frac{a^{\prime}}{a}\delta
\phi_{\mathbf{k}}^{\prime}+k^{2}\delta\phi_{\mathbf{k}}=0,
\end{equation}
where we have Fourier transformed in space and introduced the comoving
momentum $\mathbf{k}$. The conventions are such that%
\begin{equation}
\delta\phi\left(  \mathbf{x}\right)  =\frac{1}{\left(  2\pi\right)  ^{3/2}%
}\int\delta\phi_{\mathbf{k}}e^{i\mathbf{k}\cdot\mathbf{x}}d^{3}k.
\end{equation}
We have also introduced the notation $\prime$ for derivatives with respect to
conformal time. If we then introduce the rescaled field $\mu=a\delta\phi$, the
equation becomes
\begin{equation}
\mu_{\mathbf{k}}^{\prime\prime}+\left(  k^{2}-\frac{a^{\prime\prime}}%
{a}\right)  \mu_{\mathbf{k}}=0. \label{modeeq}%
\end{equation}
Similarly, the metric fluctuations can be reduced to two polarizations obeying
an equation identical to the one for the scalar fluctuations.

To proceed, treating the scalar and gravitational perturbations
simultaneously, we assume that the scale factor depend on conformal time as%
\begin{equation}
a\sim\tau^{1/2-\nu}, \label{aeta}%
\end{equation}
where $\nu$ is a constant. An important example is $a\sim e^{Ht}$ with
$H=\mathrm{const.}$, where the change of coordinates gives%
\begin{equation}
\frac{d\tau}{dt}=\frac{1}{a\left(  t\right)  }=e^{-Ht}\Longrightarrow a\left(
\tau\right)  =-\frac{1}{H\tau},
\end{equation}
and we find that $\nu=\frac{3}{2}$. Note that the physical range of $\tau$ is
$-\infty<\tau<0$. The equation for the fluctuations, with $a$ of the form
above, becomes%
\begin{equation}
\mu_{\mathbf{k}}^{\prime\prime}+\left(  k^{2}-\frac{1}{\tau^{2}}\left(
\nu^{2}-\frac{1}{4}\right)  \right)  \mu_{\mathbf{k}}=0.
\end{equation}
Luckily, this is a well known equation which is solved by Hankel functions.
The general solution is given by%
\begin{equation}
f_{k}\left(  \tau\right)  =\frac{\sqrt{-\tau\pi}}{2}\left(  C_{1}\left(
k\right)  H_{v}^{\left(  1\right)  }\left(  -k\tau\right)  +C_{2}\left(
k\right)  H_{v}^{\left(  2\right)  }\left(  -k\tau\right)  \right)  ,
\label{hank}%
\end{equation}
where $C_{1}\left(  k\right)  $ and $C_{2}\left(  k\right)  $ are to be
determined by initial conditions.

When quantizing this system (a nice treatment can be found in
\cite{Polarski:1995jg}) one needs to introduce oscillators $a_{k}\left(
\tau\right)  $ and $a_{-k}^{\dagger}\left(  \tau\right)  $ such that%

\begin{align}
\mu_{\mathbf{k}}\left(  \tau\right)   &  =\frac{1}{\sqrt{2k}}\left(
a_{k}\left(  \tau\right)  +a_{-\mathbf{k}}^{\dagger}\left(  \tau\right)
\right) \label{mupi}\\
\pi_{\mathbf{k}}\left(  \tau\right)   &  =\mu_{\mathbf{k}}^{\prime}\left(
\tau\right)  +\frac{1}{\tau}\mu_{\mathbf{k}}\left(  \tau\right)
=-i\sqrt{\frac{k}{2}}\left(  a_{k}\left(  \tau\right)  -a_{-\mathbf{k}%
}^{\dagger}\left(  \tau\right)  \right)  ,\nonumber
\end{align}
obey standard commutation relations. The crux of the matter is that these
oscillators are time dependent, and can be expressed in terms of oscillators
at a specific moment in time using the Bogolubov transformations
\begin{align}
a_{\mathbf{k}}\left(  \tau\right)   &  =u_{k}\left(  \tau\right)
a_{\mathbf{k}}\left(  \tau_{0}\right)  +v_{k}\left(  \tau\right)
a_{-\mathbf{k}}^{\dagger}\left(  \tau_{0}\right) \label{oscutv}\\
a_{-\mathbf{k}}^{\dagger}\left(  \tau\right)   &  =u_{k}^{\ast}\left(
\tau\right)  a_{-\mathbf{k}}^{\dagger}\left(  \tau_{0}\right)  +v_{k}^{\ast
}\left(  \tau\right)  a_{\mathbf{k}}\left(  \tau_{0}\right)  ,\nonumber
\end{align}
where
\begin{equation}
\left\vert u_{k}\left(  \tau\right)  \right\vert ^{2}-\left\vert v_{k}\left(
\tau\right)  \right\vert ^{2}=1.
\end{equation}
The latter equation makes sure that the canonical commutation relations are
obeyed at all times if they are obeyed at $\tau_{0}$. We can now write down
the quantum field%
\begin{equation}
\mu_{\mathbf{k}}\left(  \tau\right)  =f_{k}\left(  \tau\right)  a_{\mathbf{k}%
}\left(  \tau_{0}\right)  +f_{k}^{\ast}\left(  \tau\right)  a_{-\mathbf{k}%
}\left(  \tau_{0}\right)  ,
\end{equation}
where%
\begin{equation}
f_{k}\left(  \tau\right)  =\frac{1}{\sqrt{2k}}\left(  u_{k}\left(
\tau\right)  +v_{k}^{\ast}\left(  \tau\right)  \right)
\end{equation}
is given by (\ref{hank}).

But what are the initial conditions? The usual choice is to consider the
infinite past and choose a state annihilated by the annihilation operator,
i.e.%
\begin{equation}
a_{\mathbf{k}}\left(  \tau_{0}\right)  \left\vert 0,\tau_{0}\right\rangle =0,
\label{vakekv}%
\end{equation}
for $\tau_{0}\rightarrow-\infty$. As we will see in the next section, there is
much to say about this way to proceed, but let us, for the moment, continue
according to common practise. From (\ref{mupi}) we conclude that
\begin{equation}
\pi_{\mathbf{k}}\left(  \tau_{0}\right)  =-ik\mu_{\mathbf{k}}\left(  \tau
_{0}\right)  ,
\end{equation}
for $\tau_{0}\rightarrow-\infty$. Since the Hankel functions asymptotically
behave as%
\begin{align}
H_{v}^{\left(  1\right)  }\left(  -k\tau\right)   &  \sim\sqrt{\frac{2}%
{-k\tau\pi}}e^{-ik\tau}\nonumber\\
H_{v}^{\left(  2\right)  }\left(  -k\tau\right)   &  \sim H_{v}^{\left(
1\right)  \ast}\left(  -k\tau\right)  ,
\end{align}
we find that the vacuum choice correspond to the choice $C_{2}\left(
k\right)  =0$ (and $\left\vert C_{1}\left(  k\right)  \right\vert =1$).

We have now fully determined the quantum fluctuations, and it is time to
deduce what their effect will be on the CMBR. To do this, we compute the size
of the fluctuations according to
\begin{equation}
P\left(  k\right)  =\frac{4\pi k^{3}}{\left(  2\pi\right)  ^{3}}\left\langle
\left\vert \delta\phi_{\mathbf{k}}\right\vert ^{2}\right\rangle =\frac{k^{3}%
}{2\pi^{2}}\frac{1}{a^{2}}\left\langle \left\vert \mu_{\mathbf{k}}\right\vert
^{2}\right\rangle =\frac{k^{3}}{2\pi^{2}}\frac{1}{a^{2}}\left\vert
f_{k}\right\vert ^{2}=\frac{k^{3}}{2\pi^{2}}\frac{1}{a^{2}}\frac{\left\vert
-\tau\pi\right\vert }{4}\left\vert H_{v}^{\left(  1\right)  }\left(
-k\tau\right)  \right\vert ^{2}.
\end{equation}
This we should evaluate at late times, that is, when $\tau\rightarrow0$. In
this limit the Hankel function behaves as%
\begin{equation}
H_{v}^{\left(  1\right)  }\left(  -k\tau\right)  \sim\sqrt{\frac{2}{\pi}%
}\left(  -k\tau\right)  ^{-\nu},
\end{equation}
and we find%
\begin{equation}
P\sim\frac{1}{4\pi^{2}}\frac{1}{a^{2}}\left(  -\tau\right)  ^{1-2\nu}%
k^{3-2\nu}\sim\frac{1}{4\pi^{2}}H^{2}k^{3-2\nu}.
\end{equation}
Here we have used (\ref{aeta}) to get rid off the $\tau$ dependence.
Furthermore, if $\nu\sim3/2$ and we have a slow roll, $\ H$ is nearly constant
and can be used to set the scale of the fluctuations. In particular, we find
the well known scale invariant spectrum if $\nu=3/2$,%
\begin{equation}
P=\frac{1}{4\pi^{2}}H^{2}. \label{pt}%
\end{equation}

This is more or less the whole story in case of the gravitational, or
\textit{tensor}, perturbations. As previously explained, the scalar
fluctuations obey a similar equation, but the translation into the
perturbation spectrum is a bit more involved. Basically, different values of
$\phi$ lead to different times for the end of inflation according to $\delta
t\sim\frac{\delta\phi}{\dot{\phi}}\sim\frac{H}{2\pi\dot{\phi}}$, see, e.g.,
\cite{Linde:2002ws}. If inflation ends later, the decay of vacuum energy, and
hence the initiation of a more conventional cosmology with $H\sim1/t$ and
$\rho\sim3M_{4}^{2}H^{2}\sim1/t^{2}$, will be delayed. Therefore, we will find
an enhanced density according to $\frac{\delta\rho}{\rho}\sim\frac{\delta
t}{t}\sim\frac{H^{2}}{2\pi\dot{\phi}}$, and the relevant spectrum becomes, in
this case,%
\begin{equation}
P_{s}\sim\left(  \frac{H}{\dot{\phi}}\right)  ^{2}\frac{1}{4\pi^{2}}H^{2}.
\label{ps}%
\end{equation}
Comparing (\ref{pt}) and (\ref{ps}) we see that it is the scalar fluctuations
that play the most important role. It should be stressed that the spectra,
which we have obtained, are the \textit{primordial} ones. To obtain the actual
CMBR fluctuation spectra, including the acoustic peaks, which the primordial
spectra give rise to, requires a lot more work which is outside the scope of
this review.

To express deviations from scale invariance one introduces spectral indices
according to%
\begin{align}
n_{s}-1  &  =\frac{d\ln P_{s}}{d\ln k}=3-2\nu_{s}\\
n_{T}  &  =\frac{d\ln P_{T}}{d\ln k}=3-2\nu_{T},
\end{align}
where $\nu_{s}$ refers to the scalar perturbations and $\nu_{T}$ refers to the
gravitational, or tensor, perturbations. While not clear from our simplified
analysis, the $\nu$'s need not be the same in the two cases. Observations show
that $n_{s}$ is very close to $1$, consistent with the basic ideas of
inflation. Of extreme importance is to find any slight deviation from the
scale invariant value which could give important information about the
inflationary potential. Equally interesting would be to find a contribution
from the gravitational background.

Inflation has turned out to be a wonderful opportunity to connect the physics
of the large with physics of the small. Perhaps effects of physics beyond the
Planck scale might be visible on cosmological scales in the spectrum of the
CMBR fluctuations? This is the subject to which we now turn.

\bigskip

\subsection{Transplanckian physics}

As described in the previous section, quantum fluctuations play an important
role in the theory of inflation. But how is the structure of these microscopic
fluctuations determined? Is the standard argument that we have gone through
really valid? In a time dependent background -- where there are no global
timelike Killing vectors -- the definition of a vacuum is highly non trivial.
In the ideal situation the time dependence is only transitionary, starting out
with an initial, asymptotically Minkowsky like region, where it is possible to
one define a unique initial in-vacuum. This vacuum will time evolve through
the intermediate time dependent era, and then end up in a final Minkowsky like
region. Typically, the initial vacuum will not evolve into the final vacuum
but instead appear as an excited state with radiation. Technically, as I have
explained, one says that the excited state is related to the vacuum through a
\textit{Bogolubov transformation}. A well known example is a star that
collapses into a black hole and subsequently emits Hawking radiation.

Interestingly, a similar phenomena can be expected also during inflation. In
this case, however, the situation is more tricky since the universe (in
Robertson-Walker coordinates) is \textit{always} expanding. How can we then
choose an initial state in an unambiguous way? Luckily, the key feature of
inflation, the accelerated expansion of the universe, can help out as we have
already seen. If we follow a given fluctuation backwards in time far enough,
its wavelength will become arbitrarily smaller than the horizon radius. This
means that deviations from Minkowsky space will become less and less
important, when it comes to defining the vacuum, and the vacuum becomes, in
this way, essentially unique. This is the unique vacuum we used in the
previous section, and it is sometimes called \textit{the Bunch-Davies vacuum}.
The fact that a unique vacuum is picked out is an important property of
inflation and is one of several examples of how inflation does away with the
need to choose initial conditions.

But, and this is the main point, the argument relies on an ability to follow a
mode to infinitely small scales which, clearly, is not how it works in the
real world. After all, it is generally believed that there exists a
fundamental scale -- Planckian or stringy -- where physics could be completely
different from what we are used to, and where we have very little control of
what is happening. How does this affect the argument that the inflationary
vacuum is unique? Could there be effects of new physics which will affect the
predictions of inflation? In particular one could worry about changes in the
predictions of the CMBR fluctuations. Several groups have investigated various
ways of modifying high energy physics in order to look for such modifications,
see, e.g., [12-27].

I will not discuss the specifics of the proposals of how to modify physics
beyond the Planck scale. Instead I will take a different approach, following
\cite{Danielsson:2002kx}, and provide a typical and rather generic example of
the kind of corrections one might expect due to changes in the low energy
quantum state of the inflaton field due to the unknown high energy physics. To
proceed along this direction, we need to find out \textit{when} to impose the
initial conditions for a mode with a given (constant) comoving momentum $k$.
To do this, we use, as in the previous section, conformal time, given by
$\tau=-\frac{1}{aH}$. We note that the physical momentum $p$ and the comoving
momentum $k$ are related through
\begin{equation}
k=ap=-\frac{p}{\tau H},
\end{equation}
and impose the initial conditions when $p=\Lambda$. $\Lambda$ is the energy
scale, maybe the Planck scale or possibly the string scale, where
fundamentally new physics becomes important. The basic idea is that we do not
know what happens at higher energies, or shorter wavelengths, and therefore
are forced to encode our ignorance in terms of initial conditions when the
modes enter into the regime that we understand. The unknown high energy
physics is usually referred to as \textit{transplanckian}, and the hope is,
obviously, that, e.g., string theory eventually will give us the means to
derive these initial conditions. Proceeding with the calculation, we find the
conformal time when the initial condition is imposed to be
\begin{equation}
\tau_{0}=-\frac{\Lambda}{Hk}. \label{eta0}%
\end{equation}
As we see, different modes will be created at different times, with a smaller
linear size of the mode (larger $k$) implying a later time.

From the above it is clear that the choice of vacuum is a highly non trivial
issue in a time dependent background. Without knowledge of the transplanckian
physics we can only list various possibilities and investigate whether there
is a typical size or signature of the new effects. A useful example is to
choose the vacuum as determined by equation (\ref{vakekv}), but with an
important difference. We do \textit{not }take $\tau_{0}\rightarrow-\infty$,
but instead stop at the value of conformal time given by (\ref{eta0}). This
vacuum, which in general is different from the Bunch-Davies (note that for
$\tau_{0}\rightarrow-\infty$ the Bunch-Davies vacuum is recovered), should be
viewed as a typical representative of natural initial conditions (in the sense
explained above). It can be characterized as a vacuum corresponding to a
minimum uncertainty in the product of the field and its conjugate momentum,
\cite{Polarski:1995jg}, the vacuum with lowest energy (lower than the
Bunch-Davies) \cite{Starobinsky:2001kn}, or as the instantaneous Minkowsky
vacuum\footnote{As observed in \cite{Bozza:2003pr} the exact caracterization
of the vacuum depends on the canonical variables used.}. Therefore, it can be
argued to be as special as the Bunch Davies vacuum, and there is no a priori
reason for transplanckian physics to prefer one over the other.

We have now a one parameter family of vacua with the single parameter given by
the fundamental scale. What is the expected fluctuation power spectrum?
Following \cite{Danielsson:2002kx} one finds%

\begin{align}
P(k)  &  =\left(  \frac{H}{\overset{\cdot}{\phi}}\right)  ^{2}\left\langle
\left\vert \phi_{k}\left(  \tau\right)  \right\vert ^{2}\right\rangle =\left(
\frac{H}{\overset{\cdot}{\phi}}\right)  ^{2}\frac{1}{a^{2}}\left\langle
\left\vert \mu_{k}\left(  \tau\right)  \right\vert ^{2}\right\rangle
\label{eq:main}\\
&  =\left(  \frac{H}{\overset{\cdot}{\phi}}\right)  ^{2}\left(  \frac{H}{2\pi
}\right)  ^{2}\left(  1-\frac{H}{\Lambda}\sin\left(  \frac{2\Lambda}%
{H}\right)  \right)  ,
\end{align}
with the standard case recovered when $\Lambda\rightarrow\infty$. The result
should be viewed as a typical example of what to be expected from
transplanckian physics if we allow for effects which at low energies reduce to
modifications of the Bunch-Davies case. We note that the size of the
correction is linear in $H/\Lambda$, and that a Hubble constant, which varies
during inflation, gives rise to a modulation of the spectrum. As argued in
\ \cite{Danielsson:2002kx}, the modulation is expected to be a quite generic
effect that is present regardless of the details of the transplanckian
physics. (See also \cite{Niemeyer:2002kh} for a discussion about this). After
being created at the fundamental scale the modes oscillate a number of times
before they freeze. The number of oscillations depend on the size of the
inflationary horizon and therefore changes when $H$ changes. A varying Hubble
constant is crucial for a detectable signal, since a Hubble constant which
does not vary during inflation would just imply a small change in the overall
amplitude of the fluctuation spectrum and would not constitute a useful
signal. Luckily, since the Hubble constant \textit{is} expected to vary, the
situation is much more interesting.

Let me now turn to a more detailed discussion of possible observable
consequences. I will discuss what happens using the slow roll parameters. It
is not difficult to show (using that $H$ is to be evaluated when a given mode
crosses the horizon, $k=aH$) that
\begin{equation}
\frac{dH}{dk}=-\frac{\varepsilon H}{k},
\end{equation}
which gives
\begin{equation}
H\sim k^{-\varepsilon}.
\end{equation}
The $k$ dependence of $H$ will translate into a modulation of $P(k)$, with a
periodicity given by
\begin{equation}
\frac{\Delta k}{k}\sim\frac{\pi H}{\varepsilon\Lambda}.
\end{equation}

To be more specific, let me consider a realistic example. In the
Ho\v{r}ava-Witten model \cite{PW}, unification occurs at a scale roughly
comparable with the string scale, the higher dimensional Planck scale, as well
as the scale where the fifth dimension becomes visible. For a discussion and
references see, e.g., \cite{Polchinski:rr} or \cite{Kaloper:2002uj}. As a
rough estimate we therefore put $\Lambda=2\cdot10^{16}$ GeV -- a rather
reasonable possibility within the framework of the heterotic string. Using
that the Hubble constant during inflation can not be much larger than
$H=7\cdot10^{13}$ GeV, corresponding to $\varepsilon=0.01$, we find
\begin{align}
\frac{H}{\Lambda}  &  \sim0.004\\
\frac{\Delta k}{k}  &  =\Delta\ln k\sim1.
\end{align}
This implies one oscillation per logarithmic interval in $k$, which fits
comfortable within the parts of the spectrum covered by high-precision CMBR
observation experiments.

As I have already emphasized, it is important to note that the transplanckian
effects, regardless of their precise nature, have a rather generic signature
in form of their modulation of the spectrum. If it had just been an overall
shift or tilt of the amplitude, it would not have been possible to measure the
effect even if it had been considerably larger than the percentage level.
Instead, the only result would be a slight change in the inferred values of
$H$ and the slow roll parameters. With a definite signature, on the other
hand, we can use several measurement points throughout the spectrum, as
discussed in more detail in \cite{Bergstrom:2002yd}. There it was argued that
the upcoming Planck satellite might be able to detect transplanckian effects
at the $10^{-3}$ level, which would put the Ho\v{r}ava-Witten model within
range, or at least tantalizingly close. In this way one can also beat cosmic
variance that otherwise would have limited the sensitivity to about $10^{-2}$
at best. Other discussions can be found in \cite{Martin:2003kp}%
\cite{Elgaroy:2003gq}\cite{Martin:2003sg}\cite{Okamoto:2003wk}.

There has been extensive discussions of these results in the literature and
their relevance for detectable transplanckian signatures. As pointed out in
\cite{Danielsson:2002qh}, the initial condition approach to the transplanckian
problem allows for a discussion of many of the transplanckian effects in terms
\textit{the }$\alpha$\textit{-vacua.} These vacua have been known since a long
time, \cite{alpha}, and corresponds to a family of vacua in de Sitter space
which respects all the symmetries of the space time.

In \cite{Kaloper:2002uj}\cite{problem} concerns were raised that there could
be inconsistencies in field theories based on non trivial vacua of this sort.
None of these problems are, however, necessarily relevant to the issue of
transplanckian physics in cosmology for a very simple reason, as explained in
\cite{Ulf:0210}. The whole point with the transplanckian physics is to find
out whether effects beyond quantum field theory can be relevant for the
detailed structure of the fluctuation spectrum of the CMBR. In the real world
we do expect quantum field theory to break down at high enough energy to be
replaced by something else, presumably string theory. The modest proposal
behind \cite{Danielsson:2002kx} is simply that we should allow for an
uncertainty in our knowledge of physics near planckian scales. Several later
works, e.g., \cite{Schalm:2004qk}\cite{deBoer:2004nd}, have confirmed this
point of view and the CMBR remains a promising candidate for finding evidence
of transplanckian physics.

\bigskip

\section{String theory with and without inflation}

\bigskip

Much of contemporary cosmology has dealt with the construction of
phenomenologically viable inflationary models with various potentials and
number of inflaton fields. In the early days of inflationary theory there were
hopes of incorporating inflation in more or less standard particle physics.
Perhaps the inflaton was related to, say, the GUT-transition? Unfortunately
this never worked out in a convincing way and, as a result, inflation lived
its own life quite detached from the rest of theoretical particle physics.

Luckily, string theory is about to change all that. In string theory it is
well known that parameters describing background geometries and
compactifications, the moduli, are all promoted into scalar fields. There are,
therefore, no lack of potential candidates for the inflaton, even though there
are several difficult conditions to be met. For one thing, the potential of
the inflaton must be extremely flat in order to allow for enough e-foldings.
On the other hand, it can not be completely flat for the idea to work. In
supersymmetric string theory there are many flat directions in the moduli
space of solutions which could, it seems, serve as useful starting points. The
hope would then be that these flat directions are lifted by non perturbative,
supersymmetry breaking terms. Unfortunately, it is difficult to find these non
perturbative corrections explicitly, and their expected form is anyway, in
many cases, not of the right kind. In addition, there are also other problems
to be solved. Apart from the flat, inflationary potential, one needs
potentials that manage to fix dangerous moduli like those controlling the size
of the extra dimensions. It is hard to see how realistic inflationary theories
can be obtained without addressing this problem at the same time.

A little later I will explain some recent progress in the subject which
suggests that realistic inflationary models can indeed be constructed using
string moduli if one introduces branes. The idea is to use two stacks of
branes separated by a certain distance, corresponding to the inflaton, in a
higher dimensional space. As the branes move, the inflaton rolls, and when the
branes collide inflation stops. This is a rapidly developing subject -- for an
early review see \cite{Quevedo:2002xw}, and for more recent discussions, see
\cite{Linde:2004kg}, involving many aspects of string theory. But before
discussing these promising ideas I will discuss a couple of other interesting
approaches to cosmology.

First I will treat the attempts which go under the, somewhat unspecific, name
of \textit{string cosmology}, \cite{Veneziano:1991ek}\cite{Meissner:1991zj}%
\cite{Gasperini:1992em}\cite{Brustein:1994kw}\textit{ }(for a review see
\cite{Gasperini:2002bn}). The idea is to make use of the\textit{ dilaton},
i.e. the field corresponding to the way the string coupling varies over space
and time, and a variant of the string theoretical T-duality. The resulting
theory fulfills the condition for inflation, albeit in an unorthodox way.

After this I will turn to models based on branes. Even if branes might very
well be the key to realize inflation in string theory, they have, ironically,
also been used to argue that string theory can provide an \textit{alternative}
to inflation. I will treat a couple of such proposals, the \textit{ekpyrotic}
and the\textit{ cyclic} universe where colliding branes again play an
important role.

\subsection{String cosmology}

\bigskip

\subsubsection{The action of string cosmology}

String cosmology makes use of one of the most basic features of string theory,
the dilaton. According to string theory the Hilbert action of general
relativity is augmented by a new, dimensionless scalar field, the dilaton
$\phi$, and given by%
\begin{equation}
S=-\frac{1}{2\kappa_{10}^{2}}\int d^{10}x\sqrt{-g}e^{-\phi}\left(
\mathcal{R}+\partial^{\alpha}\phi\partial_{\alpha}\phi\right)  ,
\label{dilverk}%
\end{equation}
where $\kappa_{10}=\frac{1}{2}\left(  2\pi\right)  ^{7}\alpha^{\prime4}\sim
l_{s}^{8}$, and where the string coupling is related to the dilaton through
$g_{s}^{2}=e^{\phi}$. The action as given is written in the \textit{string
frame}. That is, the string length, $l_{s}$, is our fundamental unit and what
we use as our measuring rod. This means that the Planck mass, the effective
coefficient of the scalar curvature $\mathcal{R}$, varies with the dilaton. An
alternative way to describe things is to use the \textit{Einstein frame} which
in many ways is physically more transparent than the string frame. In the
Einstein frame it is the Planck length -- which is more directly related to
macroscopic physics through the strength of gravity -- which is used as a
fundamental unit. Let me explain how the frames are related to each other in a
little more detail. To go from one frame to another, we note that the frames
are, by definition, related through%
\begin{equation}
\int d^{D}x\sqrt{-g}e^{-\phi}\mathcal{R}=\int d^{D}x\sqrt{-g_{E}}\left(
\mathcal{R}_{E}+...\right)  ,
\end{equation}
where%
\begin{equation}
g_{\mu\nu}=e^{2\omega\phi}g_{E,\mu\nu},
\end{equation}
with the subscript $E$ indicating Einstein frame, and furthermore%
\begin{equation}
\sqrt{-g}=e^{D\omega\phi}\sqrt{-g_{E}}.
\end{equation}
It follows from the definition of curvature that the scalar curvatures are
related through%
\begin{equation}
\mathcal{R}=e^{-2\omega\phi}\left(  \mathcal{R}_{E}-2\omega\left(  D-1\right)
\nabla^{2}\phi-\omega^{2}\left(  D-2\right)  \left(  D-1\right)
\partial^{\alpha}\phi\partial_{\alpha}\phi\right)  .
\end{equation}
Hence we have that%
\begin{equation}
\sqrt{-g}e^{-\phi}\mathcal{R}=e^{\left(  D\omega-1-2\omega\right)  \phi}%
\sqrt{-g_{E}}\left(  \mathcal{R}_{E}-2\omega\left(  D-1\right)  \nabla^{2}%
\phi-\omega^{2}\left(  D-2\right)  \left(  D-1\right)  \partial^{\alpha}%
\phi\partial_{\alpha}\phi\right)  ,
\end{equation}
and as a consequence we find%
\begin{equation}
D\omega-1-2\omega=0\Longrightarrow\omega=\frac{1}{D-2}.
\end{equation}
The action in the Einstein frame finally becomes%
\begin{equation}
S=-\frac{M_{D}^{D-2}}{2}\int d^{D}x\sqrt{-g_{E}}\left(  \mathcal{R}_{E}%
-\frac{1}{D-2}\partial^{\alpha}\phi\partial_{\alpha}\phi\right)  ,
\label{Eakt}%
\end{equation}
where $M_{D}$ is the $D$-dimensional Planck mass. We note that the sign of the
kinetic term of the scalar field now is the familiar one.

If we consider a metric of FRW-form (\ref{plattmet}) generalized to $D$
dimensions, we find%
\begin{equation}
d^{2}s_{E}=e^{-2\omega\phi}d^{2}s=e^{-2\omega\phi}\left(  dt^{2}%
-a^{2}d\mathbf{x}^{2}\right)  \equiv dt_{E}^{2}-a_{E}^{2}d\mathbf{x}^{2},
\end{equation}
where%
\begin{align}
a_{E}  &  =e^{-\omega\phi}a\label{ttE}\\
dt_{E}  &  =e^{-\omega\phi}dt.\nonumber
\end{align}
It is important to realize that the two frames are physically equivalent, even
if things can, at a first glance, look rather different in the two frames. To
fully appreciate string cosmology it is important to keep this in mind.

\bigskip

\subsubsection{General idea}

\bigskip

Let us now investigate the above action in more detail. I will perform the
analysis in the string frame, and, for simplicity, assume a spatially
homogenous RW-metric. One can readily check that the scalar curvature in these
coordinates is given by%
\begin{equation}
\mathcal{R}=-\left(  D-1\right)  \left(  D-2\right)  \frac{\dot{a}^{2}}{a^{2}%
}-2\left(  D-1\right)  \frac{\ddot{a}}{a}.
\end{equation}
The action looks rather innocent, but possesses a remarkable symmetry thanks
to the presence of the stringy dilaton. The symmetry acts on the scale factor
and the dilaton through the transformations%
\begin{align}
a\left(  t\right)   &  \rightarrow1/a\left(  t\right)  \nonumber\\
\phi\left(  t\right)   &  \rightarrow\phi\left(  t\right)  -2\left(
D-1\right)  \ln a\left(  t\right)  .\label{tsymm}%
\end{align}
It leaves the action invariant and assures that the solutions of the equations
of motion have some very interesting properties that will be important for
cosmology. To verify the symmetry, we note that%
\begin{align}
\sqrt{-g}e^{-\phi}\left(  \mathcal{R}+\dot{\phi}^{2}\right)   &
=a^{D-1}e^{-\phi}\left(  -\left(  D-1\right)  \left(  D-2\right)  \frac
{\dot{a}^{2}}{a^{2}}-2\left(  D-1\right)  \frac{\ddot{a}}{a}+\dot{\phi}%
^{2}\right)  \nonumber\\
&  =a^{D-1}e^{-\phi}\left(  \left(  D-1\right)  \left(  D-2\right)  \frac
{\dot{a}^{2}}{a^{2}}-2\left(  D-1\right)  \frac{\dot{a}}{a}\dot{\phi}%
+\dot{\phi}^{2}\right)  \nonumber\\
&  +\mathrm{total\ derivative}\\
&  =a^{D-1}e^{-\phi}\left(  -\left(  D-1\right)  \frac{\dot{a}^{2}}{a^{2}%
}+\left(  \dot{\phi}-\left(  D-1\right)  \frac{\dot{a}}{a}\right)
^{2}\right)  \\
&  +\mathrm{total\ derivative}%
\end{align}
Since we have%
\begin{align}
a^{D-1}e^{-\phi} &  \rightarrow a^{-\left(  D-1\right)  }e^{-\phi+2\left(
D-1\right)  \ln a}=a^{D-1}e^{-\phi}\nonumber\\
\frac{\dot{a}}{a} &  \rightarrow a\frac{d}{dt}\left(  \frac{1}{a}\right)
=-\frac{\dot{a}}{a},
\end{align}
we find%
\begin{align}
&  a^{D-1}e^{-\phi}\left(  -\left(  D-1\right)  \frac{\dot{a}^{2}}{a^{2}%
}+\left(  \dot{\phi}-\left(  D-1\right)  \frac{\dot{a}}{a}\right)
^{2}\right)  \nonumber\\
&  \rightarrow a^{D-1}e^{-\phi}\left(  -\left(  D-1\right)  \frac{\dot{a}^{2}%
}{a^{2}}+\left(  \dot{\phi}-2\left(  D-1\right)  \frac{\dot{a}}{a}+\left(
D-1\right)  \frac{\dot{a}}{a}\right)  ^{2}\right)  ,
\end{align}
and hence an invariance of the action! In other words, if $a\left(  t\right)
$ and $\phi\left(  t\right)  $ solves the equations of motion, so does the
transformed functions $1/a\left(  t\right)  $ \ and $\phi\left(  t\right)
-2\left(  D-1\right)  \ln a\left(  t\right)  $.

To fully appreciate what is going on, and to understand the structure of the
solutions, we need to note that there is yet another simple symmetry,%
\begin{equation}
t\rightarrow-t, \label{tinv}%
\end{equation}
i.e. time reversal invariance, which together with (\ref{tsymm}) tells an
interesting story about possible cosmologies. Combining the two symmetries we
can map out how various solutions are related to each other. If we first focus
on the scale factor, we see how we from a given solution $a\left(  t\right)  $
can construct two new solutions according to%
\begin{align}
a\left(  t\right)   &  \rightarrow1/a\left(  t\right)  \qquad H\left(
t\right)  \rightarrow-H\left(  t\right) \\
a\left(  t\right)   &  \rightarrow a\left(  -t\right)  \qquad H\left(
t\right)  \rightarrow-H\left(  -t\right)  .
\end{align}
Figure 3 shows how this works.%
\begin{figure}
\begin{center}
\centering
\epsfysize=3.5cm
\leavevmode
\epsfbox{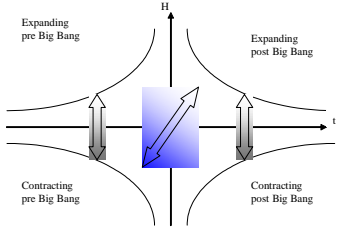}
\end{center}   
\caption[]{\small The figure shows how different cosmologies are related through the
two symmetries discussed in the text.}
\end{figure}
The time $t=0$ is referred to as the Big Bang and it is natural to allow for
two eras, a pre and a post Big Bang. The basic idea of string cosmology is
that physics can be traced back in time through the Big Bang into an earlier
era, the pre Big Bang, where many of the initial conditions for the post Big
Bang are determined in a natural and dynamical way.

It should be stressed that the whole set up is in line with the general
picture of T-duality in string theory. According to T-duality, it is
equivalent to compactify string theory on a small circle (compared with the
string scale) and a large circle. In some sense large and small scales are,
therefore, equivalent. Loosely applying this idea to the Big Bang, would
suggest that if we trace the expansion far enough back in time, we are better
off describing the universe as becoming bigger again, rather than smaller. As
we will see, however, string cosmology suggests that we should take an
\textit{expanding} pre Big Bang theory and match it to an expanding post Big
Bang. But, and this is an important but, this is the picture obtained in the
\textit{string frame}. The picture in the \textit{Einstein frame}, as I will
explain, is quite different with a contracting rather than an expanding pre
Big Bang phase. This is precisely in line with the hand waving argument above.

\bigskip

\subsubsection{An explicit example}

\bigskip

Let us now work out a detailed example to get a better feeling for how the
various cosmologies are related. In our example we add matter with a definite
equation of state,%
\begin{equation}
p=w\rho,
\end{equation}
assuming an action of the form%
\begin{equation}
S=-\frac{1}{2\kappa_{4}^{2}}\int d^{4}x\sqrt{-g}\left(  e^{-\phi}\left(
\mathcal{R}+\partial^{\alpha}\phi\partial_{\alpha}\phi\right)
+\mathrm{matter}\right)  ,
\end{equation}
with, for simplicity, no explicit $\phi$ dependence in the matter piece. We
will be using the Friedmann equations in the string frame, but, as an
exercise, we start out in the more familiar Einstein frame where the Friedmann
equations take the familiar form%
\begin{equation}
H_{E}^{2}=\frac{1}{3M_{4}^{2}}\left(  \frac{M_{4}^{2}}{2}\frac{1}{2}\left(
\frac{d\phi}{dt_{E}}\right)  ^{2}+\rho_{E}\right)  ,
\end{equation}
where we have taken the prefactor of (\ref{Eakt}) into account (with $D=4$),
when we write down the energy density for the scalar field. It is now easy,
using the relations (\ref{ttE}), to translate this into the string frame. In
particular we have%
\begin{align}
H_{E}  &  =e^{\phi/2}\left(  H-\frac{1}{2}\dot{\phi}\right) \\
\frac{d\phi}{dt_{E}}  &  =e^{\phi/2}\dot{\phi}\\
\sqrt{-g_{E}}\rho_{E}  &  =e^{-2\phi}\sqrt{-g}\rho_{E}=\sqrt{-g}\rho.
\end{align}
We finally obtain the Friedmann equation in the string frame as%
\begin{equation}
H^{2}=-\frac{1}{6}\dot{\phi}^{2}+H\dot{\phi}+\frac{1}{3M_{4}^{2}}e^{\phi}\rho.
\end{equation}

To proceed, we also need the continuum equation for matter which gives%
\begin{equation}
\rho=\rho_{0}a^{-3\left(  1+w\right)  },
\end{equation}
and the equation of motion for the dilaton obtained from the Euler-Lagrange
equation%
\begin{equation}
\frac{d}{dt}\frac{\partial L}{\partial\dot{\phi}}-\frac{\partial L}%
{\partial\phi}=0, \label{EL}%
\end{equation}
where%
\begin{equation}
L=-a^{3}e^{-\phi}\left(  -6\frac{\dot{a}^{2}}{a^{2}}+6\frac{\dot{a}}{a}%
\dot{\phi}-\dot{\phi}^{2}\right)  =6a\dot{a}^{2}e^{-\phi}-6\dot{a}a^{2}%
\dot{\phi}e^{-\phi}-a^{3}\dot{\phi}^{2}e^{-\phi}.
\end{equation}
Using an ansatz of the form%
\begin{align}
a  &  \sim t^{\alpha}\label{ansatz}\\
\phi &  =\beta\ln t+\mathrm{const.,}\nonumber
\end{align}
it is straightforward to derive, from (\ref{EL}), that%
\begin{equation}
-12\alpha^{2}-\beta^{2}+6\alpha\beta-2\beta+6\alpha=0.
\end{equation}
To fully determine $\alpha$ and $\beta$ we need one more equation. Since both
$H^{2}\sim\dot{\phi}^{2}\sim1/t^{2}$ the same must be true for $e^{\phi}\rho$
according to the Friedmann equation. The continuum equation (\ref{cont}) then
provides the missing relation%
\begin{equation}
\beta-3\left(  1+w\right)  \alpha=-2.
\end{equation}
Finally, we can write down the solution to the latter two equations as%
\begin{align}
\alpha &  =\frac{2w}{1+3w^{2}}\\
\beta &  =\frac{6w-2}{1+3w^{2}}.
\end{align}

So far we have not said what kind of matter we are considering. But let me
now, in order to be completely specific, assume that matter is in the form of
radiation with $w=1/3$. This gives%
\begin{equation}
\alpha=1/2\qquad\beta=0
\end{equation}
that is,
\begin{equation}
a\sim t^{1/2}\qquad\phi=\mathrm{const.}%
\end{equation}
In other words, we have a standard radiation dominated, and non-inflationary,
cosmology. In particular we have a decreasing Hubble constant given by%
\begin{equation}
H=\frac{\dot{a}}{a}=\frac{1}{2t}>0,
\end{equation}
with%
\begin{equation}
\ddot{a}=-\frac{1}{4t^{2}}<0\qquad\frac{\ddot{a}}{\dot{a}}<0.
\end{equation}
Not much new, but at least we see that it is consistent to have a constant
dilaton. It is now time to apply the symmetry transformations introduced
above. We immediately find a new solution given by%
\begin{equation}
a\sim\left(  -t\right)  ^{-1/2}\qquad\phi=-6\ln\left(  -t\right)
^{1/2}+\mathrm{const.,}%
\end{equation}
valid for $t<0$.\footnote{One should note that we have assumed that the matter
piece also respects the symmetries. In the particular example that we study,
this implies that the equation of state becomes $w=-1/3$ in the transformed
theory.} We now have%
\begin{equation}
H=\frac{\dot{a}}{a}=-\frac{1}{2t}>0\qquad\frac{\ddot{a}}{\dot{a}}=\frac
{3}{2t^{2}}>0
\end{equation}
with
\begin{equation}
\dot{H}=\frac{1}{2t^{2}}>0,
\end{equation}
i.e., a growing curvature. To summarize, we find an inflating universe with
growing curvature and coupling as $t\rightarrow0_{-}$, followed by a standard
radiation dominated cosmology. In other words, a rather appealing cosmology.
At least if we somehow can find a way of matching the two solutions at the Big Bang.

There is, however, another interesting twist to the story. As seen in the
previous section, the description of the physics is quite different if we
change to the Einstein frame. In our example, there is no real difference in
the post Big Bang era between the two frames, since the dilaton is constant.
In the pre-big bang phase, on the other hand, we find, using (\ref{ttE}),%
\begin{align}
dt_{E}  &  \sim\left(  -t\right)  ^{3/2}dt\qquad\Longrightarrow\qquad
t_{E}\sim-\left(  -t\right)  ^{5/2}\\
a_{E}  &  \sim\left(  -t\right)  ^{3/2}\times\left(  -t\right)  ^{-1/2}%
=-t\sim\left(  -t_{E}\right)  ^{2/5},
\end{align}
and as a consequence%
\begin{align}
a_{E}^{\prime}  &  \sim-\left(  -t_{E}\right)  ^{-3/5}<0\\
a_{E}^{\prime\prime}  &  \sim-\left(  -t_{E}\right)  ^{-8/5}<0.
\end{align}
The physical picture in the Einstein frame is therefore of a contracting
rather than an expanding universe. Nevertheless we can rest assured that the
physics will be equivalent.

In order to understand better what is going on, it is useful, following
\cite{Gasperini:2002bn}, to classify the various possibilities according to
the following table:

\medskip%

\begin{tabular}
[c]{|l|l|l|l|}\hline
Class I &  &  & \\\hline
$\dot{a}>0$ & $\ddot{a}>0$ & $\dot{H}<0$ & standard inflation\\\hline
Class II &  &  & \\\hline
$\dot{a}>0$ & $\ddot{a}>0$ & $\dot{H}>0$ & superinflation\\\hline
$\dot{a}<0$ & $\ddot{a}<0$ & $\dot{H}<0$ & collapse!\\\hline
\end{tabular}

\medskip

As in our example, superinflation and a collapsing universe can be different
descriptions of the same physics in string and Einstein frames
respectively.\footnote{One should note that superinflation driven by a
standard scalar field is not possible in the Einstein frame. This will be
discussed in a different context a bit later.} In figure 4 one can see how
this works.
\begin{figure}
\begin{center}
\centering
\epsfysize=3.5cm
\leavevmode
\epsfbox{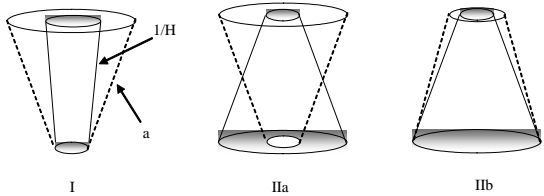}
\end{center}   
\caption[]{\small Three ways of fulfilling the inflationary criterium.}
\end{figure}
It is interesting to see that the advantages of inflation can be obtained also
in a contracting universe. The important thing is that the ratio of the radius
of curvature and the scale factor becomes smaller with time.

\bigskip

\subsubsection{Problems of string cosmology}

As I have already hinted, a basic problem of string cosmology is how to match
the pre and post Big Bang solutions. This is known as\textit{ the graceful
exit problem.} As is clear from the examples above, the matching has to take
place at strong coupling and little is known about how to achieve this. I will
come back to the same problem in the next section, when I discuss some
alternative models.

Another important issue is the CMBR-fluctuations. Let me continue to discuss
the particular example introduced above. To apply the formulae of section
3.1.1. we need to go to conformal time. We find%
\begin{equation}
\tau\sim-\left(  -t_{E}\right)  ^{3/5}%
\end{equation}
and%
\begin{equation}
a_{E}\sim\left(  -\tau\right)  ^{2/3}\sim\left(  -\tau\right)  ^{1/2-\nu_{T}%
}\qquad\Longrightarrow\qquad\nu_{T}=-1/6,
\end{equation}
and from this%
\begin{equation}
n_{T}=\frac{10}{3}.
\end{equation}
That is, a blue spectrum for the gravitational perturbations not at all like
the more or less scale invariant result of standard inflation. This is
certainly an interesting prediction and could be a characteristic signal to
look for if, and when, these perturbations become observationally accessible.
Unfortunately, however, a similar spectrum can be derived also for the scalar
fluctuations which dominate the CMBR. This is not at all in line with what
observations show, and is one of the big problems with the simplest approaches
to string cosmology. Some possible ways out of this dilemma is discussed in
\cite{Gasperini:2002bn}.

\subsection{Brane cosmology}

\bigskip

\subsubsection{Alternatives to inflation}

\bigskip

\paragraph{Basic setup}

\bigskip

In the middle 90's, it was realized that not only strings but also higher
dimensional structures like membranes etc. play an important role in string
theory. Moreover, branes provide new possibilities to construct realistic
cosmologies. Of particular interest is the idea to associate the Big Bang with
a collision of brane worlds which I will discuss in some detail. \ This has
been considered from two quite different points of view -- either as an
alternative to inflation or as a way of implementing inflation.

The first of the alternatives to inflation is the \textit{ekpyrotic scenario},
\cite{Khoury:2001wf}\cite{Kallosh:2001ai}\cite{Khoury:2001bz}. It makes use of
the Ho\u{r}ava-Witten interpretation of the heterotic $E_{8}\times E_{8}$
string where there is an eleventh dimension separating two 9+1 dimensional
brane worlds. The separation between the branes gives the string coupling in
such a way that a small separation corresponds to weak coupling. We are
assumed to be living on one of the branes, the visible brane, while the other
brane is called the hidden brane. In the ekpyrotic scenario there is an
additional brane in the bulk which is free to move. The configuration is
assumed to be nearly supersymmetric, i.e. BPS, and therefore nearly stable --
apart from a small potential which provides an attraction between the bulk
brane and the visible brane.

The main idea behind the ekpyrotic scenario is to let the Big Bang correspond
to a collision between the bulk brane and the visible brane. The homogeneity
of the early universe, usually explained by inflation, is explained by the
nearly BPS initial state. The bulk brane is almost parallel with the visible
brane and the collision happens almost at the same time everywhere. From the
point of view of physics on the visible brane, the era before the collision is
a contracting universe, while the era after the collision (or Big Bang) is our
expanding universe. Slight differences in collision time give rise to the
crucial primordial spectrum of fluctuations. This represents a new mechanism,
fundamentally different from the one of inflation.

An improved proposal is the \textit{cyclic scenario}, \cite{Steinhardt:2001vw}%
, where one does away with the bulk brane and lets, instead, the visible and
hidden branes collide. Actually, the branes are supposed to be able to pass
through each other and, eventually, turn back for yet another collision. And
so on, forever. The homogeneity is, in this model, explained not through
initial conditions, but by a late time cosmological constant in each cycle.
The cosmological constant provides an accelerated expansion that sweeps the
universe clean of disturbances preparing it for a new cycle. The idea is that
we presently are entering into such an era and, in this way, the model
suggests an interesting role for the cosmological constant recently observed.
In a way the cyclic universe make use of inflation of a kind, even though the
energy scales involved are totally different. Note, however, that the quantum
fluctuations during the inflationary stage in the cyclic universe will be
irrelevant for the CMBR fluctuations due to the low energy scale.

Whatever description all of this has from the point of view of higher
dimensions, there should also be an effective four dimensional picture. To
study this, we start with the same action as in string cosmology,
(\ref{dilverk}), but think of $\phi$ as a scalar field such that $e^{\phi}$ is
proportional to the distance between the branes. The Big Crunch occurs when
the distance between the branes vanish, that is when $\phi\rightarrow-\infty$,
corresponding to a Big Crunch at weak coupling since, from the four
dimensional point of view, $e^{\phi}$ is like a coupling. Note that this is
just the opposite to what we have in string cosmology. We use the same ansatz
as before, (\ref{ansatz}), in the Friedmann equation for an empty universe%
\begin{equation}
H^{2}=-\frac{1}{6}\dot{\phi}^{2}+H\dot{\phi},
\end{equation}
to get%
\begin{equation}
\alpha^{2}=-\frac{1}{6}\beta^{2}+\alpha\beta.
\end{equation}
The two equations are solved by%
\begin{equation}
\alpha=\pm\frac{1}{\sqrt{3}}\qquad\beta=\pm\sqrt{3}-1,
\end{equation}
that is,%
\begin{equation}
a\sim t^{\varepsilon/\sqrt{3}}\qquad\phi=\left(  \varepsilon\sqrt{3}-1\right)
\ln t, \label{ekpsol}%
\end{equation}
where $\varepsilon=\pm1$. If we had been doing string cosmology we would have
applied the duality transformations of (\ref{tsymm}) (and (\ref{tinv})). This
leads to%
\begin{align}
a  &  \sim\left(  -t\right)  ^{-\varepsilon/\sqrt{3}}\\
\phi &  =\left(  \varepsilon\sqrt{3}-1\right)  \ln\left(  -t\right)
-6\ln\left(  -t\right)  ^{\varepsilon/\sqrt{3}}=\left(  -\varepsilon\sqrt
{3}-1\right)  \ln\left(  -t\right)  .
\end{align}
Clearly, this is essentially an exchange of the two solutions in
(\ref{ekpsol}). In string cosmology we would have made the choice
\begin{equation}%
\begin{array}
[c]{cc}%
t<0 & \varepsilon=-1\\
t>0 & \varepsilon=+1,
\end{array}
\end{equation}
with $t\rightarrow0_{-}\Longrightarrow g_{s}\rightarrow+\infty$ and
$t\rightarrow0_{+}\Longrightarrow g_{s}\rightarrow0$. In the ekpyrotic
universe, however, where the collision of branes corresponds to weak coupling,
we have $\varepsilon=+1$ for \textit{all} $t$!

To proceed, we note that for all $t$ and $\varepsilon$, we have that%
\begin{equation}
\frac{\dot{a}}{a}\sim\frac{\varepsilon}{\sqrt{3}t}.
\end{equation}
Using this we find for string cosmology%
\begin{align}
&
\begin{array}
[c]{ccc}%
t<0 & \varepsilon=-1 & \frac{\dot{a}}{a}>0
\end{array}
\nonumber\\
&
\begin{array}
[c]{ccc}%
t>0 & \varepsilon=+1 & \frac{\dot{a}}{a}>0,
\end{array}
\end{align}
while the ekpyrotic universe has%
\begin{align}
&
\begin{array}
[c]{ccc}%
t<0 & \varepsilon=+1 & \frac{\dot{a}}{a}<0
\end{array}
\nonumber\\
&
\begin{array}
[c]{ccc}%
t>0 & \varepsilon=+1 & \frac{\dot{a}}{a}>0.
\end{array}
\end{align}
This was all in the string frame. In the Einstein frame we simply find%
\begin{align}
a_{E}  &  \sim\left(  -t_{E}\right)  ^{1/3}\qquad\mathrm{for\qquad}t_{E}<0\\
a_{E}  &  \sim t_{E}^{1/3}\qquad\mathrm{for\qquad}t_{E}>0
\end{align}
if we follow the recipe provided earlier.\footnote{This corresponds to a
universe filled by matter with equation of state given by $p=\rho$. This is
precisely what one gets from a massless scalar field without potential.} That
is, regardless of whether we are considering string cosmology or the cyclic
universe, we find a universe that first collapses and then expands. The
difference is the behavior of the scalar field. One notes that the condition
$\frac{\ddot{a}_{E}}{\dot{a}_{E}}>0$ is fulfilled in the $t_{E}<0$ era both
for string cosmology and for the cyclic universe. In fact, the process with
fluctuations crossing the horizon, and entering in a much later era, is common
to standard inflation and the ekpyrotic/cyclic universe.

\paragraph{Can it work?}

\bigskip

The ekpyrotic/cyclic scenarios have been heavily criticized in the literature,
see, e.g., \cite{Linde:2002ws}\cite{Kallosh:2001ai}. I will briefly review
some of this criticism. But let me begin by considering the generation of
fluctuations in the ekpyrotic/cyclic universe. To do this, we make use of
(\ref{sceq}), which we expand to quadratic order to get%
\begin{equation}
\delta\ddot{\phi}+3H\delta\dot{\phi}-\nabla^{2}\delta\phi+V^{\prime\prime
}\left(  0\right)  \delta\phi=0.
\end{equation}
In conformal time, assuming spatial homogeneity, we find%
\begin{equation}
\mu_{k}^{\prime\prime}+\left(  k^{2}-\frac{a^{\prime\prime}}{a}\right)
\mu_{k}+a^{2}V^{\prime\prime}\left(  0\right)  \mu_{k}=0.
\end{equation}
Usually, the last term is ignored due to the flatness of the potential -- a
necessary condition for inflation. In the cyclic scenario, however, this is no
longer the case. Instead, it is the term due to the expansion/contraction of
the universe that should be ignored. The generation of fluctuations takes
place when the universe is contracting very slowly, and the scale factor is
more or less constant. A useful potential, with the correct properties, is%
\begin{equation}
V\left(  \phi\right)  =-V_{0}e^{-\phi/M_{4}},
\end{equation}
and we will look for a solutions with $a=\mathrm{const}$. We then need to
solve%
\begin{align}
\ddot{\phi}+V^{\prime}\left(  \phi\right)   &  =0\nonumber\\
H^{2}  &  =\frac{1}{3M_{4}^{2}}\left(  \frac{1}{2}\dot{\phi}^{2}+V\right)  =0
\end{align}
It is easy to verify that this works for%
\begin{equation}
\dot{\phi}=\frac{2M_{4}}{t},
\end{equation}
and we find%
\begin{equation}
\mu_{\mathbf{k}}^{\prime\prime}+\left(  k^{2}-\frac{V_{0}}{M_{4}^{2}}%
e^{-\phi/M_{4}}\right)  \mu_{\mathbf{k}}=0,
\end{equation}
or%
\begin{equation}
\mu_{\mathbf{k}}^{\prime\prime}+\left(  k^{2}-\frac{2}{t^{2}}\right)
\mu_{\mathbf{k}}=0,
\end{equation}
which is precisely the same equation as derived in the context of inflation!
However, there are some important differences. In the inflationary case, we
need to rescale $\mu$ with the scale factor to get the original scalar field
$\phi=\mu/a$. As we have already seen, the amplitude of the fluctuations in
$\phi$ are always finite. In the ekpyrotic/cyclic case, however, the scale
factor is essentially constant and we are stuck with the field $\mu$ whose
amplitude \textit{diverges} as $t\rightarrow0$. We therefore need a cutoff
near the moment of collision. This is a reflection of the tachyonic nature of
the potential with $V^{\prime\prime}\left(  0\right)  <0$. In inflation the
classical perturbations are smeared thanks to the exponential expansion. In
the ekpyrotic universe, however, thsi does not happen and the classical
perturbations are amplified in just the same way as the quantum mechanical
ones, \cite{Linde:2002ws}\cite{Kallosh:2001ai}. Hence the need for fine tuning
of initial conditions.

In case of the cyclic universe we must also investigate the claim that the
cycles continue forever. \ A well known argument against an eternal, cyclic
universe, comes from the second law of thermodynamics. With every cycle the
entropy should increase and one would not expect an infinite number of cycles.
In case of the brane based cyclic universe, it is argued that the exponential
expansion due to the late time cosmological constant does the job through a
rapid clean up which effectively provides an empty universe ready for the next cycle.

However, it is hard to see how this statement can be \textit{exactly} true.
From the point of view of a local observer it is true that any matter
(carrying entropy) is heavily redshifted and pushed towards the cosmological
horizon. But, as I will discuss in the chapter on holography, there is a limit
on how much entropy can be stored by the horizon. When this limit is reached,
there will be unavoidable consequences for the physics of the cyclic universe.
As a result, the second law will eventually prevail after all. It is true,
though, that the time scale for this to happen will be enormous.

Another crucial problem of the ekpyrotic/cyclic proposal is the bounce. Will
the branes bounce off each other or will there be a devastating singularity?
Unfortunately, it is well known that the necessary reversal from contraction
to expansion is very difficult, if not impossible, to achieve. What is needed,
is a Hubble constant which starts out negative and then becomes positive. In
other words, we need a period with $\dot{H}>0$. The problem is that we have
the Friedmann equation%
\begin{equation}
\dot{H}=-\frac{1}{2M_{4}^{2}}\left(  \rho+p\right)  ,
\end{equation}
with a right hand side which for all reasonable types of matter is negative.
An example is the scalar field in section 2.3., where the equations
(\ref{dphi}) and (\ref{pphi}) \ yield%
\begin{equation}
\dot{H}=-\frac{1}{2M_{4}^{2}}\dot{\phi}^{2}<0.
\end{equation}
The same problem is also present in case of string cosmology, but in that case
we at least can blame strong coupling and hope that, somehow, there is a way
out. In the ekpyrotic scenario, everything happens at weak coupling suggesting
that there is little chance of evading the contradiction.

There are also other arguments indicating that a singularity is the end and
not a new beginning. The idea is that the creation of a new universe beyond
the singularity inside of a black hole, would imply that black holes are
information sinks as first suggested by Hawking. However, it is now generally
believed that string theory predicts that all information \textit{is} getting
back out from a black hole through the Hawking radiation. From this it is
argued in \cite{Horowitz:2003he} that no information can pass through a
singularity into a new baby universe. The kind of bounce needed for the
ekpyrotic or cyclic universe would therefore not take place. Very recently,
\cite{Turok:2004gb}, new arguments have been put forward where it is suggested
that a cosmological singularity \textit{can }be resolved in M-theory. It is
fair to say, therefore, that there is no consensus in the field at the moment.

What, then, is the conclusion? The ekpyrotic universe represents a different
paradigm without inflation where, instead, it is argued that high energy
physics can naturally provide very special initial conditions all on its own.
The cyclic universe does not do away with inflation completely but, in a very
economical way, identifies inflation with the presence of a cosmological
constant late in each cycle. Unfortunately, both scenarios face severe
technical problems due to the difficulties in understanding the bounce.
Whether or not string theory allows for a world beyond a time like singularity
is of crucial importance, not only to cosmology.

\subsubsection{Inflation from branes}

\bigskip

Can inflation be realized using branes? As we have seen above the distance
between two branes can be identified with a scalar field on the branes
yielding interesting cosmologies. But instead of using this to construct an
alternative to inflation, we will now try to identify the scalar field as the inflaton.

The first attempts to construct brane inflation used two sets of branes. If
the configuration preserves supersymmetry there is no force between the branes
and no potential for the scalar field. What happens is that there is a balance
between the gravitational attraction between the branes and a repulsion due to
the RR-charges of the branes. If supersymmetry is broken, however, the dilaton
and the RR-fields obtain masses while the graviton remain massless. Hence the
attraction wins and there is a force between the branes. In principle this
could yield inflation if the resulting potential is of just the right form,
\cite{Dvali:1998pa}. Unfortunately, our understanding of string theory is not
deep enough to enable us to perform trustworthy calculations with
nonperturbative supersymmetry breaking. Actually, the situation is not unlike
what we have in the ekpyrotic/cyclic scenario where the actual potential also
is not very well known.

Another possibility, in the sense that we can perform reliable calculations to
check the scenario, is to consider brane and anti-branes where supersymmetry
is broken and there is a force already at tree level. In this case the branes
have opposite charges and there is an attractive RR-force that adds to the
gravitational force. Let us see how this works, \cite{Burgess:2001fx}%
\cite{Shiu:2001sy}\cite{Kyae:2001mk}.

We start out with the action of a Dp-brane in $10$ dimensional space time. It
is given by%
\begin{equation}
S_{D}=-\int d^{4}xd^{p-3}y\sqrt{-h}\left[  T_{p}+...\right]  ,
\end{equation}
where
\begin{equation}
T_{p}=\frac{1}{\left(  2\pi\right)  ^{p}}\alpha^{\prime-\frac{p+1}{2}}%
e^{-\phi}\sim\frac{1}{g_{s}l_{s}^{p+1}}, \label{spann}%
\end{equation}
is the tension of the brane, $h_{\mu\nu}$ is the metric on the branes induced
by the full metric $g_{\mu\nu}$, and $x$ corresponds to our four space time
coordinates while $y$ are compact extra dimensions around which the D-branes
are wrapped (if $p>3$). The action for the \={D}p-brane is identical. The
position of the D-brane and the \={D}-brane in the transverse dimensions are
denoted by $z_{1}^{m}$ and $z_{2}^{m}$ respectively, where $m=1...d_{\perp}$
with $d_{\perp}=10-\left(  p+1\right)  =9-p$ as the number of transverse
dimensions. We now define the relative position of the D and \={D}-branes as
$z^{m}=z_{1}^{m}-z_{2}^{m}$, and their average position as $\bar{z}^{m}%
=\frac{z_{1}^{m}+z_{2}^{m}}{2}$. The two actions added together, can then be
expanded as%
\begin{equation}
S_{D}+S_{\bar{D}}=-\int d^{4}xd^{p-3}y\sqrt{-\bar{h}}T_{p}\left[  2+\frac
{1}{4}g_{mn}\bar{h}^{\mu\nu}\partial_{\mu}z^{m}\partial_{\nu}z^{n}+...\right]
, \label{branverk}%
\end{equation}
where $\bar{h}_{\mu\nu}$ is evaluated on $\bar{z}^{m}$. We now have the
kinetic terms for our inflaton field $z$, but what about the potential?%

\begin{figure}
\begin{center}
\centering
\epsfysize=3.5cm
\leavevmode
\epsfbox{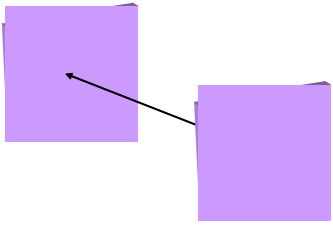}
\end{center}   
\caption[]{\small Two D-branes with their distance interpreted as the inflaton.}
\end{figure}

The potential energy is of the same form as the gravitational potential
between two branes, that is, the energy per area is given by%
\begin{equation}
\frac{E}{A_{p}}=-\beta\frac{1}{M_{10}^{2}}\frac{T_{p}^{2}}{z^{d_{\perp}-2}},
\end{equation}
where $\beta=\frac{1}{8}\pi^{-d_{\perp}/2}\Gamma\left(  \frac{d_{\perp}-2}%
{2}\right)  $, and $M_{10}^{2}=e^{-2\phi}\kappa_{10}^{-2}$. Compactifying
according to%
\begin{equation}
\frac{M_{10}^{2}}{2}\int d^{4}xd^{6}y\sqrt{-g}\left[  R+...\right]  \sim
\frac{M_{4}^{2}}{2}\int d^{4}x\sqrt{-g}\left[  R+...\right]  ,
\end{equation}
we find%

\begin{equation}
M_{4}^{2}=M_{10}^{2}V_{\perp}V_{\parallel}. \label{mpvv}%
\end{equation}
The six extra dimensions should be compact, and we assume that they have
volumes given by%
\begin{equation}
V_{\perp}\equiv r_{\perp}^{d_{\perp}},\qquad V_{\parallel}\equiv r_{\parallel
}^{p-3}.
\end{equation}
The potential (including the mass density of the branes) can, after
compactification, be written%
\begin{equation}
V\left(  z\right)  =2T_{p}V_{\parallel}-\frac{\beta}{M_{10}^{2}}\frac
{T_{p}^{2}V_{\parallel}}{z^{d_{\perp}-2}}\equiv A-\frac{B}{z^{d_{\perp}-2}}.
\end{equation}
To complete the calculation, we also need to make sure that our inflaton has
the correct normalization. Looking at (\ref{branverk}) we see that we need to
identify the canonically normalized scalar field as%
\begin{equation}
\psi=\sqrt{\frac{T_{p}V_{\parallel}}{2}}z.
\end{equation}
Can the resulting potential yield inflation? To answer this question, we need
to evaluate the slow roll parameters. These are given by%
\begin{align}
\varepsilon &  =\frac{M_{4}^{2}}{2}\left(  \frac{V^{\prime}}{V}\right)
^{2}\sim\frac{M_{4}^{2}}{T_{p}V_{\parallel}}\left(  \frac{B}{A}\left(
d_{\perp}-2\right)  \frac{1}{z^{d_{\perp}-1}}\right)  ^{2}\nonumber\\
&  =\frac{\beta^{2}}{4}\left(  d_{\perp}-2\right)  \frac{T_{p}}{M_{10}^{2}%
}\frac{1}{z^{d_{\perp}-2}}\left(  \frac{r_{\perp}}{z}\right)  ^{d_{\perp}}\sim
g_{s}\left(  \frac{l_{s}}{z}\right)  ^{d_{\perp}-2}\left(  \frac{r_{\perp}}%
{z}\right)  ^{d_{\perp}}\\
\eta &  =M_{4}^{2}\frac{V^{\prime\prime}}{V}\sim-\frac{2M_{4}^{2}}%
{T_{p}V_{\parallel}}\frac{B}{A}\left(  d_{\perp}-2\right)  \left(
d_{\perp}-1\right)  \frac{1}{z^{d_{\perp}}}\nonumber\\
&  =-\beta\left(  d_{\perp}-2\right)  \left(  d_{\perp}-1\right)  \left(
\frac{r_{\perp}}{z}\right)  ^{d_{\perp}}\sim\left(  \frac{r_{\perp}}%
{z}\right)  ^{d_{\perp}},
\end{align}
where we have made use of (\ref{mpvv}). The derivatives of $V$ are taken with
respect to $\psi$. From the requirement that $\eta$ should be small, we
immediately see that $z\gg r_{\perp\text{ }}$which, unfortunately, does not
make sense. The branes can not be separated by a distance larger than the size
of the compact dimension!

One possible way out, is to fine tune the positions of the D and \={D} to
opposite sides of the compact dimension. From symmetry this must correspond to
a meta-stable, forceless configuration. It can be shown that the potential
close to the equilibrium position is such that slow roll and inflation is
allowed. In the next section we will come back to other possibilities of
obtaining realistic models.

It is also interesting to think about what will happen when the branes
collide. From string theory we would expect the annihilation of the branes to
be driven, from the perspective of the brane, by an open string tachyon. The
field $T$ corresponding to the tachyon becomes tachyonic when the distance
between the branes is decreased to a string length, \cite{Banks:1995ch}. We
therefore expect a potential of the form%
\begin{equation}
V\left(  z,T\right)  =A\left(  \frac{z^{2}}{l_{s}^{2}}-B\right)  T^{2}%
+CT^{4}+V\left(  z\right)  ,
\end{equation}
where $A,B$ and $C$ are positive constants. Interestingly, this is just the
kind of potential known from \textit{hybrid inflation}, \cite{Linde:1993cn}%
.\textit{ }The original motivation for hybrid inflation was to generate enough
e-foldings without the inflaton having to start out with values of the order
of many Planck masses, recall (\ref{n70}). Contrary to a single field chaotic
inflation, where all of the vacuum energy decays through the rolling inflaton,
the decay in hybrid inflation takes place in two steps, as shown in figure 6.%
\begin{figure}
\begin{center}
\centering
\epsfysize=3.5cm
\leavevmode
\epsfbox{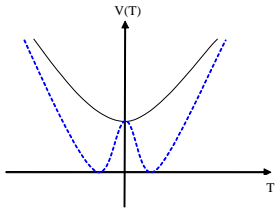}
\end{center}   
\caption[]{\small The tachyon potential for large values of $z$, and for small values of
$z$ (dashed line).}
\end{figure}
First, when $z$ is large, the tachyon $T$ is locked in a minimum at $T=0$. The
effective potential for the inflaton can, in the simplest case, be of the
usual monomial type, but the minimum has a non-zero vacuum energy that can
drive inflation. However, when $z$ becomes small enough, $T=0$ becomes
unstable and rolls down to a new minimum. As a result, the vacuum energy
decays away. In brane inflation, this corresponds to the annihilation of the branes.

Unfortunately we do not have a good understanding of what happens when the
branes annihilate and how rehetaing takes place. That is, how all the matter
now present in the universe is created out of the decaying vacuum energy. We
must also make sure that all branes do not annihilate after the collision.
There must be a net number of, say D-branes, remaining after all pairs of D
and \={D} have annihilated.

\subsection{Strings in de Sitter space}

\bigskip

Recent observations give very strong indications that we are living in a
universe with a positive cosmological constant, i.e. a de Sitter space. From
the point of view of string theory this is quite surprising. In fact, it has
been a long standing problem to formulate string theory in de Sitter space.
Part of the difficulty has to do with supersymmetry. Contrary to the case of
flat space time and a space time with a negative cosmological constant, i.e.
anti de Sitter space, a positive cosmological constant goes together with
super symmetry breaking. It has, therefore, not been possible to take
advantage of the simplifications due to supersymmetry in constructing de
Sitter space times. Another, more serious problem, is that string theory is
naturally formulated using S-matrices. That is, we need an asymptotic
lightlike infinity, like in Minkowsky space, to make sense of the scattering
amplitudes produced by string theory. An exception is anti de Sitter space,
where we have the option to describe physics holographically on the time like
boundary. Unfortunately, neither of these possibilities are available in de
Sitter space. Based on this, it was argued in \cite{Witten:2001kn} that an
accelerated expansion, like the one due to a cosmological constant,
necessarily is temporary.

A seemingly different problem in string theory is the stabilization of moduli.
For instance, why is the size of the compact dimensions stable? Why do they
not change in a substantial way during the evolution of the universe? As we
will see in the following these two problems are not unrelated to each other.

Let us start out, following \cite{Giddings:2003zw}, with the action
\begin{equation}
S=\int d^{D}x\sqrt{-g}\left[  -\frac{M_{D}^{D-2}}{2}\mathcal{R}+L\left(
\psi\right)  \right]  ,
\end{equation}
with $D=d+4$, and metric%
\begin{equation}
ds^{2}=ds_{4}^{2}+R^{2}\left(  x\right)  g_{dmn}\left(  y\right)  dy^{m}%
dy^{n}.
\end{equation}
We dimensionally reduce to four dimensions and find%
\begin{equation}
S=-\frac{M_{D}^{D-2}}{2}V_{d}\int d^{4}x\sqrt{-g_{4}}e^{d\phi\left(  x\right)
}\mathcal{R}_{4}+V_{d}\int d^{4}x\sqrt{-g_{4}}e^{d\phi\left(  x\right)  }\int
d^{d}y\sqrt{g_{d}}L\left(  \psi\right)  ,
\end{equation}
where we have put%
\begin{align}
R\left(  x\right)   &  =R_{0}e^{\phi\left(  x\right)  }\\
V_{d}  &  =R_{0}^{d}.
\end{align}
Now let us rescale%
\begin{equation}
g_{4,\mu\nu}\rightarrow e^{-d\phi\left(  x\right)  }g_{4,\mu\nu}.
\end{equation}
This leads to%
\begin{align}
\sqrt{-g_{4}}  &  \rightarrow e^{-2d\phi\left(  x\right)  }\sqrt{-g_{4}}\\
\mathcal{R}_{4}  &  \rightarrow e^{d\phi\left(  x\right)  }\mathcal{R}_{4}+...
\end{align}
and the action becomes%
\begin{equation}
S=-\frac{M_{4}^{2}}{2}\int d^{4}x\sqrt{-g_{4}}\mathcal{R}_{4}+V_{d}\int
d^{4}x\sqrt{-g_{4}}e^{-d\phi\left(  x\right)  }\int d^{d}y\sqrt{g_{d}}L\left(
\psi\right)  .
\end{equation}
It is important to note that the second term includes a factor $e^{-d\phi
\left(  x\right)  }$ that decreases if the volume of the compact dimensions
increases. This can only be compensated if $\int d^{d}y\sqrt{g_{d}}L\left(
\psi\right)  $ goes like the volume, i.e. like $R^{d}$. The metric $g_{d}$
does not include any $R$-dependence which leave us with the density $L\left(
\psi\right)  $. There is no type of matter which has an energy density that
grows with the volume of space. In fact, recalling (\ref{rhoa}) we see that
such matter would have $w=-2$. The best we can do is to consider cases where
there is effectively a cosmological constant. This can be obtained by wrapping
a brane around the compact dimensions.

In general, we find that the energy approaches zero as the dimensions
decompactify, and we end up, eventually in ten dimensional flat space time.
There are several possibilities for how this can happen, one of which is
illustrated in figure 7.%
\begin{figure}
\begin{center}
\centering
\epsfysize=3.5cm
\leavevmode
\epsfbox{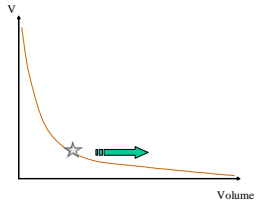}
\end{center}   
\caption[]{\small Energy is gained by letting the extra dimensions decompactify.}
\end{figure}
In this case there is nothing that can prevent the compact dimensions from
opening up, and the system rapidly rolls towards the decompactified case.
Another possibility, illustrated in figure 8,
\begin{figure}
\begin{center}
\centering
\epsfysize=3.5cm
\leavevmode
\epsfbox{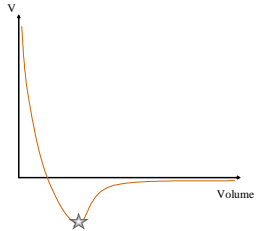}
\end{center}   
\caption[]{\small Anti de Sitter space time corresponds to a possibility of having
stable compact extra dimensions.}
\end{figure}
is that there is a minimum of the potential at negative energy, i.e., an anti
de Sitter universe where the compact dimensions are stabilized. The much
studied $AdS^{5}\times S^{5}$ is an example of this. Finally, there could be a
(local) minimum with positive energy corresponding to a de Sitter universe as
shown in figure 9.
\begin{figure}
\begin{center}
\centering
\epsfysize=3.5cm
\leavevmode
\epsfbox{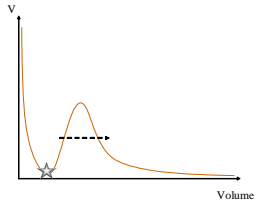}
\end{center}   
\caption[]{\small A de Sitter space time with a positive cosmological constant is not
excluded, but is unstable against tunneling.}
\end{figure}
The size of the extra dimensions are now meta stable -- eventually there will
be a tunneling to the decompactified case.

Interestingly, the cases of figure 8 can be realized in string theory. In
\cite{Giddings:2001yu}\cite{DeWolfe:2002nn} type IIB with six dimensions
compactified on Calabi-Yau spaces were studied. By turning on fluxes, the
complex moduli of the internal spaces were stabilized, and in
\cite{Kachru:2003aw} it was noted that non perturbative string corrections can
also fix the volume of the internal space. Hence we end up in a situation
described by figure 8. \ Then, the authors of \cite{Kachru:2003aw} added a
number of $\bar{D}3$ branes, which increased the energy and corrected the
potential to the one in figure 8.

Interestingly, the model also provides a way of realizing brane inflation. The
trick is to make use of the fact that the $\bar{D}3$ branes are sitting at
fixed positions on the internal manifold at the bottom of deep throats. If we
add some $D3$ branes these will move down the throats attracted by the
$\bar{D}3$ branes. Thanks to the redshift at the bottom of the throats, the
problem of achieving slow roll in a compact dimension that I discussed
previously is circumvented, \cite{Kachru:2003sx}.

It remains to construct realistic models within string theory that provide the
right amount of inflation, the correct cosmological constant of today, as well
as realistic particle physics. But the indications are certainly there that it
should be possible. Are there any generic predictions? Most of the D-brane
based string models discussed in the recent literature has an inflationary
scale that is rather low. This means that $\epsilon$ is essentially zero and
any deviation from scale invariance comes from $\eta$. From an observational
point of view this is slightly disappointing for two reasons. First, a too
small $\varepsilon$ implies that contributions to the CMBR fluctuations from
gravitational waves will be non-observable. Second, the magnitude of
transplanckian effects in the CMBR, in the simplest and most generic
scenarios, will be beyond detection. It is therefore of great interest to find
out whether an almost vanishing $\varepsilon$ is a robust prediction of string theory.

\section{Holography}

\bigskip

Holography is an intriguing possibility for finding connections between the
smallest scales and cosmology. I will not give a review of all the various
attempts to apply holography to cosmology. Some of the more original and
interesting are discussed in \cite{holo}. Instead, I will describe a number of
important and general features of holography that I find important to keep in
mind. The subject is, unfortunately, full of contrary claims and confusions,
and my aim is to put the subject on as solid ground as possible.

I will start out with a discussion of entropy bounds and the question of
whether such bounds can provide useful restrictions on cosmology, not
available by other means. My conclusion will be negative. Then I will proceed
with a discussion of more intricate questions like complementarity. Here, the
answer is not as clear cut, but my conclusion will, nevertheless, be that
there is no known mechanism for how such effects could be made visible.

\subsection{Holographic bounds}

\bigskip

Holography has its origin in black hole physics and the discovery in the 70's
by Bekenstein that black holes carry an entropy proportional to the area of
the horizon, \cite{Bekenstein:ax}. Bekenstein further argued that there are
general bounds on the amount of entropy that can be contained in matter. The
entropy bound, due to Bekenstein, that will serve as a starting point for my
discussion states that in asymptotically flat space, \cite{Bekenstein:jp}, is
\begin{equation}
S\leq S_{B}=2\pi ER,
\end{equation}
where $E$ is the energy contained in a volume with radius $R$. This is the
\textit{Bekenstein bound}. There are several arguments in support of the bound
when gravity is weak \cite{info}, and it is widely believed to hold true for
all reasonable physical systems. Furthermore, in the case of a black hole
where $R=2El_{pl}^{2}$, we have an entropy given by
\begin{equation}
S_{BH}=\frac{A}{4l_{pl}^{2}}=\frac{\pi R^{2}}{l_{pl}^{2}},
\end{equation}
which exactly saturates the Bekenstein bound. We will consequently put
$\hbar=c=1$, but explicitly write the Planck length, $\ l_{pl}=\sqrt
{\frac{G\hbar}{c^{3}}}$, to keep track of effects due to gravity.

Beginning with \cite{Fischler:1998st}, there have been many attempts to apply
similar entropy bounds to cosmology and in particular to inflation,
\cite{holo}. The idea has been to choose an appropriate volume and argue that
the entropy contained within the volume must be limited by the area. An
obvious problem in a cosmological setting is, however, that for a constant
energy density a bound of this type will always be violated if the radius $R$
of the volume is chosen to be big enough. In fact, this observation has been
used to argue, choosing appropriate volumes, that holography puts meaningful
limits on, e.g., inflation. However, as was explained in \cite{hubble}, it is
not reasonable to discuss radii which are larger than the Hubble radius in the
expanding universe. See also \cite{Brustein}. This, then, suggests that the
maximum entropy in a volume of radius $R>r$, where $r$ is the Hubble radius,
is obtained by filling the volume with as many Hubble volumes as one can fit
-- all with a maximum entropy of $\frac{\pi r^{2}}{l_{pl}^{2}}$. This gives
rise to the \textit{Hubble bound}, which states that
\begin{equation}
S<S_{H}\sim\frac{R^{3}}{r^{3}}\frac{r^{2}}{l_{pl}^{2}}=\frac{R^{3}}%
{rl_{pl}^{2}}.
\end{equation}
The introduction of the Hubble bound removes many of the initial confusions in
the subject of holographic cosmology.

\bigskip

\subsection{Local versus global perspectives}

\bigskip

The Hubble bound \ is a bound on the entropy that can be contained in a volume
much larger than the Hubble radius. It is, therefore, a bound that gives
measurable consequences only if inflation stops allowing scales larger than
the inflationary Hubble radius to become visible. Clearly, the notion of a
cosmological horizon, and its corresponding area, does not play an important
role from this point of view.

If we, on the other hand, want to discuss things from the point of view of
what a local observer, who do not have time to wait for inflation to end, can
measure, we must be more careful. In this case one has a cosmological horizon
with an area that it is natural to give an entropic interpretation
\cite{GibbHawk:1977}. Since the area of the horizon grows when matter is
passing out towards the horizon, from the point of view of the local observer,
it is natural to expect the horizon to encode information about matter that,
in its own reference frame, has passed to the \textit{outside} of the
cosmological horizon of the local observer. From the point of view of the
observer, the matter will never be seen to leave but rather become more and
more redshifted. The outside of the cosmological horizon should, therefore, be
compared with the inside of a black hole. It follows that the horizon only
indirectly provides bounds on entropy within the horizon as is nicely
exemplified through the \textit{D-bound} introduced in \cite{Bousso:2000md}.
The cosmological horizon area in a de Sitter space with some extra matter is
smaller than the horizon area in empty space. If the matter passes out through
the horizon, the increase in area can be used to limit the entropy content in
matter. This is the content of the D-bound which turns out to coincide with
the Bekenstein bound. The D-bound, therefore has not, necessarily, that much
to with de Sitter space or cosmology. It is more a way to use de Sitter space
to derive a constraint on matter itself.

Let me now explain the nature and relations between the various entropy bounds
a little bit better. In particular on what scales the entropy is stored. If we
assume that all entropy is stored on short scales smaller than the horizon
scale $r$, we can consider each of the horizon bubbles separately and use the
Bekenstein bound (or D-bound) on each and everyone of these volumes. We
conclude from this that the entropy, under the condition that it is present
only on small scales, is limited by
\begin{equation}
S<S_{LB}=2\pi Er,
\end{equation}
which I will refer to as the \textit{local Bekenstein bound}. It is
interesting to compare this result with the entropy of a gas in thermal
equilibrium. One then finds $S_{g}\lesssim Er$ for high temperatures where
$T\gtrsim1/r$, and $S_{g}\gtrsim Er$ for low temperatures where $T\lesssim
1/r$. This is quite natural and a consequence of the fact that most of the
entropy in the gas is stored in wavelengths of the order of $1/T$. This means
that the entropy for low temperatures is stored mostly in modes larger than
the Hubble scale and can therefore violate the local Bekenstein bound $S_{LB}$.

The size of the horizon therefore limits the amount of information on scales
larger than the Hubble scale, or, more precisely, the large scale information
that once was accessible to the observer on small scales. If the horizon is
smaller than its maximal value this is a sign that there is matter on small
scales, and the difference limits the entropy (or information) stored in the
matter. This is the role of the D-bound. We conclude, then, that a system with
an entropy in excess of $S_{LB}$ (but necessarily below $S_{H}$) must include
entropy on scales larger than the horizon scale.

While the entropy bounds above are rather easy to understand, the way entropy
can flow and change involve some more subtle issues. In the case of a diluting
gas the expansion of the universe implies a flow of entropy out through the
horizon, but as the gas eventually is completely diluted the flow of entropy
taps off. Whether or not the horizon radius is changing, one will never be
able to violate the Hubble bound or get an entropy flow through an apparent
horizon violating the bound set by the area. A potentially more disturbing
situation is obtained if we consider an empty universe (apart from a possibly
changing cosmological constant), which can be traced arbitrarily far back in
time, with entropy generated through the quantum fluctuations that are of
importance for the CMBR. As discussed in several works, \cite{Polarski:1995jg}%
\cite{squeeze}, there is an entropy production that can be associated with
these fluctuations and one can worry that this will imply an entropy flow out
through the horizon that eventually will exceed the bound set by the horizon.
This is the essence of the argument put forward in \cite{Albrecht:2002xs}.

To understand this better, one must have a more detailed understanding of the
cause of the entropy. Entropy is always due to some kind of coarse graining
where information is neglected. In the case of the inflationary quantum
fluctuations we typically imagine, as I have explained, that the field starts
out in a pure state -- defined by some possibly transplackian physics -- with
a subsequent unitary evolution that keeps the state pure for all times. This
is true whether we take the point of view of a local observer or use the
global FRW-coordinates. To find an entropy we must introduce a notion of
coarse graining. Various ways of coarse graining have been proposed, but they
all imply an entropy that grows as the state gets more and more squeezed,
\cite{Polarski:1995jg}\cite{squeeze}. It can be shown that most of this
entropy is produced at large scales (when the modes are larger than the
horizon), and well below the Hubble bound.

This is all in terms of the FRW-coordinates, but let us now take the point of
view of the local observer. In this case the freedom to coarse grain is more
limited. In order to generate entropy we must divide the system into two
subsystems and trace out over one of the subsystems in order to generate
entropy in the other. As an example consider a system with $N$ degrees of
freedom divided into two subsystems with $N_{1}$ and $N_{2}$ degrees of
freedom, respectively, with $N=N_{1}+N_{2}$ and $N_{2}>N_{1}$. If the total
system is in a pure state it is easy to show that the entropy in the larger
subsystem is limited by the number of degrees of freedom in the smaller one,
i.e. $S_{2}<\ln N_{1}$.\footnote{A simple proof can be found in
\cite{Danielsson:um} in the context of the black hole information paradox.}
Applied to our case, this means that the entropy flow towards the horizon must
be balanced by other matter with a corresponding ability to carry entropy
within the horizon. Since the amount of such matter is limited by the D-bound,
the corresponding entropy flow is also limited. As a consequence, there can
not be an accumulated flow of entropy out towards the horizon that is larger
than the area of the horizon. For a similar conclusion see
\cite{Frolov:2002va}. This does not mean that inflation can not go on for
ever, nor that there can not be a steady production of entropy on large
scales, but it does imply that the local observer will not be able to do an
arbitrary amount of coarse graining.

To summarize: \textit{from a local point of view the production of entropy in
quantum fluctuations is limited by the ability to coarse grain; from a global
point of view entropy is created on scales larger than the Hubble scale.}

\bigskip

\subsection{Complementarity}

\bigskip

I have argued that holography, in the sense of putting limits on the entropy,
does not constrain cosmology in any new way. It might still be a useful
principle, but it does not contain anything beyond what is contained in the
Bekenstein bound and the generalized second law which, in turn, seem to be
automatically obeyed by the ordinary laws of physics. If we want to find truly
new effects, we must go one step further and turn to the principle of
\textit{complementarity}. I will therefore investigate the possibilities of an
\textit{information paradox} and compare with the corresponding situation in
the case of black holes.

In black hole physics, the emerging view is that a kind of complementarity
principle is at work implying that two observers, one travelling into a black
hole and the other remaining on the outside, have very different views of what
is going on. According to the observer staying behind, the black hole explorer
will experience temperatures approaching the Planck scale close to the
horizon, and as a consequence, the black hole explorer will be completely
evaporated and all information transferred into Hawking radiation. According
to the explorer herself, however, nothing peculiar happens as she crosses the
horizon. As explained in \cite{Susskind}, the apparent paradox is resolved
when one realizes that the two observers can never meet again to compare
notes. Any attempts of the observers to communicate again, after the outside
observer have extracted the information from the Hawking radiation, will
necessarily make use of planckian energies and presumably fail.

An interesting question to pose is whether a similar mechanism could be at
work also in de Sitter space. In order to investigate such a possibility, we
will consider a scenario where at some moment in time the de Sitter phase is
turned off and replaced by a non-accelerated $\Lambda=0$ phase with ordinary
matter. That is, an inflationary toy model. A possible information paradox,
comes about if one assumes that an object receding towards the de Sitter
horizon of an inertial de Sitter observer, will return its information content
to the observer in the form of de Sitter radiation. If the cosmological
constant turns off, the object itself will eventually return to the observers
causal patch, and one has the threat of a duplication of information and
therefore a paradox.

To come to terms with the paradox, let us focus on what an observer actually
would see as an object recedes towards the horizon, \cite{Danielsson:2003wb}.
Since the rate of the photons (emerging from the horizon) received by our
observer is of order $1/R$, the time it would take for her to see the object
burn will be extremely long. To find out how long, we will investigate what
actually happens to the object (according to the observer). To do that we
think of the horizon as an area consisting of $R^{2}/l_{pl}^{2}$ Planck cells,
and remember that the photon has a wavelength of order Planck scale when
emitted and can indeed resolve specific Planck cells.

Now, let us assume the object in question to be something really simple, with
an information content much smaller than the $R^{2}$ number of degrees of
freedom of the horizon. This would mean that only a few of the Planck cells
are involved in encoding the object. In the extreme case of an object with
entropy of $\mathcal{O}(1)$, one would need to wait until of the order $R^{2}$
photons have been emitted to be reasonably sure to see a photon coming from
the burning of the object. In the other extreme, one can think of an object
consisting of the order $R^{2}$ degrees of freedom. In this case it is clear
that one has to wait until of the order $R^{2}$ photons have been emitted, in
order for all parts of the object to have been burnt. Regardless of the size
of the object, one has, therefore, to wait a time,
\begin{equation}
\tau\sim\frac{R^{2}}{l_{pl}^{2}}R=\frac{R^{3}}{l_{pl}^{2}}\sim\frac{1}%
{T^{3}l_{pl}^{2}},
\end{equation}
in order to actually see the destruction.

If we now abruptly turn off the de Sitter phase, and let it be followed by a
more standard cosmological evolution, we expect the object to eventually
return to the observers causal patch at some time in the future
\cite{Danielsson:2002td}. If the time we wait before turning of inflation is
shorter then the estimate above, this causes no problem. The object simply
becomes visible again with a negligible amount of de Sitter radiation emitted.
If we wait longer the situation is more confusing. The time we have estimated
is the time it takes for an object to be irreversibly lost to the horizon, and
it would be inconsistent for an unharmed, information loaded object to come
back. After all, we have, with our own eyes, seen the object burn. This is, in
fact, just the information paradox.

Luckily, the time scale we have estimated above is long enough for several
interesting effects to take place which have the potential of removing the
paradox. One argument goes as follows. Since the situation relation between
the object and the observer is symmetric, it is clear that the object will be
in as good, or bad, shape as the observer. Indeed, considering the symmetric
situation we have between the observer and the object (being for example
another observer) and the fact that they can meet again some time after the de
Sitter phase has turned off, seems to imply that the estimated time should be
the same for local objects as for those who approach the horizon, even from
the perspective of one single observer.

So, let us now try to estimate the time it takes to break down a local object,
bound to the observer. To do this, we reconsider the possibility that local
interactions do give rise to a breakdown, but only if we take physics near the
Planck scale into account. With an interaction rate given by $\Gamma=\sigma
nv$, where the cross section is given by $\sigma\sim l_{pl}^{2}$, the number
density of the radiation $n\sim T^{3}\sim1/R^{3}$ and the relative velocity
$v=c=1$, one finds the typical time $\tau$ it takes for this process to occur
to be
\begin{equation}
1\sim\sigma nv\tau\sim l_{pl}^{2}\cdot1/R^{3}\cdot1\cdot\tau\mbox
{ }\Rightarrow\mbox{ }\tau\sim\frac{R^{3}}{l_{pl}^{2}}.
\end{equation}
This coincides, up to orders of one, with the previous result. Therefore,
regardless of whether local objects or objects falling towards the horizon are
concerned, the survival time will be the same. We argued above that this must
be the case based on the symmetry between the observer and the object and by
noting that, if the de Sitter phase is only temporary, they will eventually
meet again. We find it encouraging that the above results are in agreement
with this assessment.

The above analysis provides a possible escape route from the information
paradox, since, as I have argued, it is very difficult for an observer to
exist long enough to actually see any object being fully burnt by Hawking
radiation. But this is not all, as observed in \cite{Danielsson:2002td} there
is a further obstacle to experiencing an information paradox. It can be shown
that the return time for an object that has been falling towards the horizon a
time $\tau\sim\frac{R^{3}}{l_{pl}^{2}}$, is of the order of the Poincare
recurrence time, $\sim e^{R^{2}/l_{pl}^{2}}$ of the de Sitter space. That is,
it exceeds the Poincare recurrence time of the detector.

What are the implications for inflation? In inflation the Hubble constant is
constrained from observations to be no larger than $H\sim10^{-4}\mbox{ }M_{4}%
$. With this input the thermalization time for non-thermal excitations
($\alpha$-vacua included) is found to be of order $\tau\sim R^{3}=1/H^{3}%
\sim10^{12}\mbox{ }t_{pl}$. Comparing this with the time needed for the
required number of e-foldings, which for 70 e-foldings is $t_{infl}%
\sim70/H\sim7\cdot10^{5}\mbox{ }t_{pl}$, one concludes that the thermalization
time allows for visible effects of non-thermal behavior in the CMBR, with room
to spare. This is good news for the transplanckian signatures. On the other
hand, with fluctuations leaving the horizon so close to the end of inflation,
effects from holography and complementarity are expected to be subtle.

A fair conclusion is to say that so far holography has not yielded any useful
restrictions on cosmology.

\bigskip

\section{Conclusions}

\bigskip

Is there a unique theory? String theory was long described as an attempt to
find the unique theory of the world. The idea was that mathematical necessity
fully determines the fundamental laws of nature leaving no room for chance or
choice. While well in line with the previous success story of fundamental
physics, there is something unnerving with this idea. Many scientists have
pointed out that the fundamental laws of physics are very well adjusted to
create a universe suitable to life. It has been argued that a small change of
any of the 20 or so parameters of the standard model, would leave us with a
barren and inhospitable universe. How can it be that mathematics cares about life?

With this in mind it is reasonable to question whether the fundamental laws
really are unique. It seems more reasonable to imagine that the universe is
much larger than what we so far have had reason to believe. In different
regions the fundamental parameters take different values and we, obviously,
live in a region where everything is just right. This is the anthropic principle.

Ironically, recent developments in string theory suggests that this scenario
could be correct. The hope of finding one unique string theory is far from
being realized. Instead, there seems to be an infinite number of consistent
compactifications in various dimensions, \cite{Susskind:2003kw}. In addition,
inflation provides a mechanism for generating all possible vacua,
\cite{Linde:2002gj}.

In these lectures I have discussed a few possibilities of using cosmological
observations to test string theory. In particular I have argued that physics
near the Planck scale might leave an imprint on the CMBR and give the first
glimpses of how Nature works on its smallest scales. I have reviewed attempts
to build realistic models of cosmology using strings and branes. Finally, I
have also discussed holography and the open question of whether issues like
entropy bounds and complementarity could be of importance for, in particular, inflation.

So far, our theoretical understanding of string theory and quantum gravity in
a cosmological setting is rudimentary -- a lot of work certainly remains to be
done. But from what we already know, it is reasonable to hope that we one day
will see the effects of strings in the sky.

\section*{Acknowledgments}

The author would like to thank Lars Bergstr\"{o}m and Martin Olsson for
fruitful collaborations. The author is a Royal Swedish Academy of Sciences
Research Fellow supported by a grant from the Knut and Alice Wallenberg
Foundation. The work was also supported by the Swedish Research Council (VR).

\end{document}